\documentclass{article}

\usepackage{arxiv}

\usepackage[utf8]{inputenc} % allow utf-8 input
\usepackage[T1]{fontenc}    % use 8-bit T1 fonts
\usepackage{hyperref}       % hyperlinks
\usepackage{url}            % simple URL typesetting
\usepackage{booktabs}       % professional-quality tables
\usepackage{amsfonts}       % blackboard math symbols
\usepackage{nicefrac}       % compact symbols for 1/2, etc.
\usepackage{microtype}      % microtypography
\usepackage{lipsum,enumerate,bm,mathrsfs,yfonts}
\usepackage{graphicx,amsmath,multicol,multirow}
\usepackage{url}
\usepackage{amsthm}
\usepackage[style=apa, backend=biber, natbib=true]{biblatex}
\usepackage{adjustbox,xr,comment} 

\usepackage{amsmath}
\usepackage{amssymb}
\usepackage{amsthm,tikz}
\usepackage{colortbl,caption,xr}
\usepackage{adjustbox}  
\usepackage{graphicx}
\usepackage{lscape}
\usepackage{rotating}
\usepackage{graphicx}
\usepackage{float}
\usepackage{url}
\usepackage{subfigure}
\usepackage{dcolumn}
\usepackage{multirow}
\usepackage{verbatim,booktabs, array,arydshln}
\usepackage{color}
\usepackage{setspace}
\usepackage{bbm}
\usepackage{booktabs}
\usepackage{hhline}
\usepackage{bm}
\usepackage{dsfont,import}
\usepackage{enumerate}
\usepackage{threeparttable}
\usepackage{array}
\usepackage{caption}
\usepackage{tikz}
\usetikzlibrary{shapes.geometric, arrows, shadows}
\usepackage{graphicx}

\usepackage{siunitx}

\newtheorem{theorem}{Theorem}[section]   % numbered within sections
\theoremstyle{definition}

\theoremstyle{remark}
\newtheorem*{remark}{Remark}

\theoremstyle{assumption}
\newtheorem{assumption}[theorem]{Assumption}

\theoremstyle{proposition}
\newtheorem{proposition}[theorem]{Proposition}

\tikzstyle{state} = [ellipse, draw, fill=gray!20, text centered, text width=1.5cm, minimum height=1.2cm, drop shadow]
\tikzstyle{arrow} = [thick,->,>=stealth]

\title{Doubly robust estimators of the restricted mean time in favor estimands in individual- and cluster-randomized trials}

\author{
 Xi Fang \\
    Department of Biostatistics \\
    Yale School of Public Health \\
    New Haven, CT, USA\\
\And
Bingkai Wang \\
Department of Biostatistics \\
School of Public Health \\
University of Michigan \\
Ann Arbor, MI, USA\\
  \And
Guangyu Tong\\
Department of Biostatistics \\
Yale School of Public Health \\
New Haven, CT, USA\\
 \And 
Liangyuan Hu\\
Department of Biostatistics and Epidemiology\\
Rutgers School of Public Health\\
Piscataway, NJ, USA\\
\And
Shuangge Ma\\
Department of Biostatistics \\
Yale School of Public Health \\
New Haven, CT, USA\\
\And
 Fan Li \\
Department of Biostatistics \\
Yale School of Public Health \\
New Haven, CT, USA\\
  \texttt{fan.f.li@yale.edu} \\
}

\begin{document}
\maketitle
\begin{abstract}
Progressive multi-state survival outcomes are common in trials with recurrent or sequential events and require treatment effect estimands that remain interpretable without proportional intensity or Markov assumptions. The restricted mean time in favor of treatment (RMT-IF) extends the restricted mean survival time to ordered multi-state processes and provides such an interpretable estimand. However, existing RMT-IF methods are nonparametric, assume covariate-independent censoring for independent observations, and do not accommodate cluster-randomized trials (CRTs), limiting both efficiency and applicability. We develop a class of doubly robust estimators for RMT-IF under right censoring using an augmented inverse-probability weighting framework that combines stage-specific outcome regression with arm-specific censoring models, yielding consistency when either nuisance model is correctly specified. We further extend the framework to CRTs by formalizing both cluster-level and individual-level average RMT-IF estimands to address informative cluster size and by constructing corresponding doubly robust estimators that account for within-cluster correlation. For inference, we employ model-agnostic jackknife variance estimators in both individually randomized and cluster-randomized settings. Extensive simulation studies demonstrate finite-sample performance, and the methods are illustrated using two randomized trial examples.
\end{abstract}

% keywords can be removed
\keywords{Covariate-dependent censoring; covariate adjustment; doubly-robust estimation; multi-state survival outcome; restricted mean time in favor; estimands}

\section{Introduction}\label{intro}

\begin{figure}[htbp]
    \centering
    \begin{tikzpicture}[node distance=3cm]

    % Nodes
    \node (S0) [state] {state 0};
    \node (S1) [state, right of=S0] {state 1};
    \node (S2) [state, right of=S1] {state 2};
    \node (S3) [state, below of=S1, yshift=-0.5cm] {state 3};
    
    % Arrows
    \draw [arrow] (S0) -- node[anchor=south] {} (S1);
    \draw [arrow] (S0) -- node[anchor=east] {} (S3);
    \draw [arrow] (S1) -- node[anchor=south] {} (S2);
    \draw [arrow] (S1) -- node[anchor=west] {} (S3);
    \draw [arrow] (S2) -- node[anchor=west] {} (S3);
    
    \end{tikzpicture}
    \caption{An example diagram for a 3 transition-state process; state 0 can be considered as an initial state of being alive and event free.}
    \label{fig:multi_state}
\end{figure}
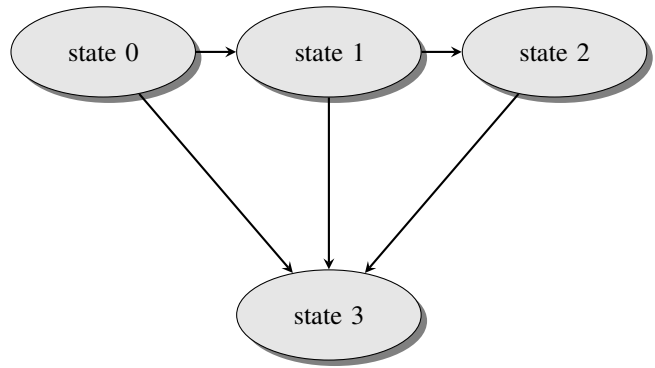

%% p1: multi-state survival outcomes
In longitudinal clinical studies, researchers often track individuals over time to observe the occurrence and timing of various events. Traditional survival analysis methods, such as the Kaplan-Meier estimator and the Cox proportional hazards model \cite{kaplan1958nonparametric,cox1972regression}, are widely used when the focus is on a single endpoint. However, these methods are limited in scope when multiple event types or sequential health states are of interest. In such scenarios, multi-state models provide a comprehensive framework for capturing the complexity of event histories. \cite{andersen2002multi} These models characterize the evolution of an individual's status as transitions among a finite set of discrete states over continuous time. In medical applications, states may correspond to clinically defined stages of the illness or treatment phases, such as cancer progression or stages of HIV infection. Each state represents the individual's condition at a given time, and movements between states are referred to as transitions that may indicate disease progression. States may be classified as transient, allowing future transitions, or absorbing, indicating finality with no further changes. For instance, in a semi-competing risks setting, once a terminal event occurs, no subsequent transitions are possible. The structure and complexity of a multi-state model depend on both the number of states and the allowable transitions between them. A basic example is the standard survival model, which consists of a transition from an initial alive state to a final death state. By decomposing the alive state into several intermediate stages, one can conceptualize a progressive multi-state framework that reflects more complex clinical pathways, as illustrated in Figure \ref{fig:multi_state}.

%% p2: estimands for survival outcomes and multi-state survival outcomes

While multi-state models provide a powerful framework for capturing the complexity of disease progression and treatment response over time, their use also raises important questions about how to define and interpret treatment effect estimands. In particular, specifying meaningful estimands for treatment comparisons becomes substantially more challenging as the model structure grows more complex. Conventional treatment effect estimands, such as those based on the restricted mean survival time (RMST) and survival probability are well-suited for simple two-state data structure involving a single transition. \cite{fay2024causal} These estimands also possess unambiguous causal interpretations under the potential outcomes framework. In contrast, the causal interpretation of commonly reported quantities such as the hazard ratio or transitional hazard ratio depends the proportional hazards assumption. In multi-state settings involving recurrent events or semi-competing risks, summarizing the treatment effect with a single, one-dimensional estimand becomes more nuanced. %Several approaches have emerged to address this complexity. 
Marginal modeling approaches have been proposed to describe the process dynamics across multiple states. For instance, Andersen et al. \cite{andersen2019modeling} and Hu et al.\ \cite{hu2011analysis} developed intensity-based models to analyze recurrent events (each recurrent event can be considered as a distinct state) in the presence of a terminal event. While such models offer valuable insights into the progression dynamics, the resulting intensity-based estimands are conditional on post-randomization events, complicating their interpretation in a potential outcomes framework. As Andersen and Keiding \cite{andersen2012interpretability} noted, drawing causal conclusions from such estimands requires careful consideration of the underlying assumptions and the marginal process features being compared. Alternatively, others have defined causal estimands based on the expected number of unfavorable events among the principal strata of always-survivors, \cite{comment2025survivor,ohnishi2025principal} or while-alive estimands that summarize the generalized while-alive loss rate of all component events.\cite{wei2023properties,fang2025while} A final class of alternative estimands have been proposed that are based on functionals of the time-to-event distribution. For example, Buhler et al.\ \cite{buhler2023multistate} considered quantiles and restricted expected times free of specific events as interpretable marginal summaries. Mao \cite{mao2023restricted} further developed the restricted mean time in favor (RMT-IF) of treatment estimand, which integrates RMST within the pairwise comparison framework. The decomposition of the RMT-IF estimand also offers a lens for state-wise evaluation of treatment effects across multiple outcome components, providing a more nuanced and clinically interpretable perspective than traditional single-outcome estimands.

%% p3: Existing methods for rmt-if estimands and limitation.
The RMT-IF estimands can be expressed as functionals of marginal state-specific survival functions defined over a sequence of landmark transition events, making them amenable to nonparametric plug-in estimation. Leveraging this foundational idea, Mao \cite{mao2023restricted} developed a two-sample inferential framework in which overall and stage-wise RMT-IF are obtained by evaluating these functionals at Kaplan-Meier estimators of the relevant landmark time distributions, with large sample inference derived from the corresponding influence function under completely independent censoring within treatment arms. However, this approach does not fully leverage baseline covariates to address covariate-dependent censoring schemes nor to improve estimation efficiency. In settings where the censoring mechanism depends on prognostic covariates, plug-in estimators based solely on marginal Kaplan-Meier curves may incur bias, and even under independent censoring, the omission of covariate adjustment can lead to avoidable efficiency loss. Although regression-type generalizations of RMT-IF have been suggested conceptually, such as replacing marginal survival functions with covariate-specific estimators based on proportional hazards models or proposing generalized regression formulations, no fully specified estimating equations or implementable procedures have been developed or numerically studied.\cite{mao2023restricted}
%Although regression-type extensions of RMT-IF have been suggested conceptually in Mao,\cite{mao2023restricted} including replacing the marginal survival estimators with covariate-specific versions based on proportional hazard models, and developing a generalized regression model for RMT-IF, there are no fully specified estimating equations or implementable procedures that have been established. \cite{mao2023restricted} 
Moreover, existing RMT-IF estimators assume independent data and therefore do not accommodate the cluster-correlated data inherent to cluster-randomized trials. In these settings, treatment is assigned at the cluster level and outcomes among individuals within the same cluster may be correlated due to shared environments, care processes, or provider behaviors. Recent work on estimands for cluster-randomized designs has emphasized that CRTs require careful articulation of the target population, as individual-average and cluster-average effect estimands may diverge in the presence of informative cluster size, and the choice between them has implications for interpretation and inference.\cite{kahan2023estimands,kahan2023demystifying,li2025model} At the same time, multi-state endpoints are becoming more common in CRTs, especially in pragmatic clinical research where participants may experience a sequence of clinically relevant events. In such contexts, RMT-IF provides a flexible framework for summarizing cumulative treatment effects across these transitions. 
Given these considerations, a clear methodological gap remains in further developing RMT-IF with independent data, as well as extending RMT-IF to cluster-randomized settings through covariate-adjusted estimators that appropriately target both individual- and cluster-average versions of the RMT-IF estimands.

%% p4: our contribution, doubly robust estimator for rmt-if in i.i.d and cluster-randomized trial. 
In this article, we introduce a covariate-adjusted doubly robust estimation framework for estimating RMT-IF that can incorporate baseline covariates without compromising the target estimand, with applications to both individual-randomized trial (IRTs) and CRTs. Our framework explicitly targets a marginal causal estimand and provides estimand-aligned inference by combining inverse probability weighting and outcome regression based on familiar survival models for the multistate outcome and censoring process (e.g., Cox regression for time to each state and censoring time). The proposed estimators are consistent when either the outcome or censoring models are correctly specified, but not necessarily both. Our approach is designed to accommodate progressive multi-state structures without relying on parametric transition models or restrictive Markov assumptions. It is also designed to accommodate covariate-dependent censoring, a common challenge that is not directly addressed by the existing two-sample nonparametric estimator.\cite{mao2023restricted} In the context of CRTs, we further extend the doubly robust estimators to account for hierarchical dependence and informative cluster size.\cite{kahan2023estimands} For valid inference, we employ efficient resampling-based variance estimators: a leave-one-group-out jackknife for IRTs and a leave-one-cluster-out jackknife for CRTs, the latter of which accommodates within-cluster correlation without requiring strong distributional assumptions. We assess the finite-sample performance of the proposed estimators through extensive simulations under varying model misspecification scenarios. All proposed methods are implemented in the \texttt{DRsurvCRT} \textsf{R} package. %supporting convenient applications across both IRT and CRT settings.

%% p4 section organization

The remainder of this article is organized as follows. In Section \ref{estimand}, we begin with the setting of IRTs and formulate the RMT-IF estimand using potential outcomes. We introduce a doubly robust estimator for this estimand in Section \ref{dr_estimator}. In Section \ref{crt_estimand}, we extend the framework to CRTs, where we formalize two distinct estimands: the cluster-average and individual-average analogues of RMT-IF. We present the extended doubly robust estimators that allow for informative cluster size. In Section \ref{jk_var_iid} and  \ref{jk_var_crt}, we describe jackknife variance estimators for each design. We present our simulation studies in Sections \ref{sec:sim}, and compare our new estimators with existing RMT-IF estimators, when applicable. Illustrative analyses of the Systolic Blood Pressure Intervention Trial (SPRINT) and the Strategies to Reduce Injuries and Develop Confidence in Elders (STRIDE) trial are presented in Section \ref{data_ex}, demonstrating the implementation of our methods in a real-world setting. Finally, Section \ref{discussion} concludes with a summary of our contributions and discussions of future research directions.

\section{A doubly robust estimator for RMT-IF in individual-randomized trials}
%Formalizing the treatment effect estimands in IRTs with multi-state survival outcomes} \label{estimand}

\subsection{Notation, setup and a review of RMT-IF framework}\label{estimand}
We consider an IRT with \( N \) individuals. Each individual \( i = 1,\dots,N \) is randomized to receive treatment \( A_i \in \{0,1\} \), where \( A_i = 1 \) denotes receipt of the active intervention and \( A_i = 0 \) denotes usual care. Each individual is also characterized by a vector of baseline covariates \( \bm{Z}_i \in \mathbb{R}^p \). Over continuous time, individuals may transition through a series of clinically meaningful states that represent increasing levels of severity, culminating in a terminal absorbing state such as death. We define the potential multistate outcome process \( Y_i^{(a)}(t) \in \{0,1,\dots,Q+1\} \), which denotes the health state of individual \( i \) at time \( t \) had they been assigned to treatment \( a \in \{0,1\} \). The value \(0,1,\dots,Q+1\) are used solely as state labels and do not imply equal spacing between states. The state space is assumed to be ordered, with a larger value corresponding to a worse health status, and \( Q+1 \) representing a terminal absorbing state. For pairwise comparisons, we consider a generic pair of potential trajectories $Y^{(1)}(t)$ and $Y^{(0)}(t)$ sampled independently from their respective treatment arms. By convention, we assume all individuals begin in the initial healthy state 0 at baseline. The evolution of the process \( Y_i^{(a)}(t) \) is assumed to follow a progressive structure, meaning individuals can only move to equal or worse states over time, and cannot revert to less severe states, as stated in the following. %This progression is captured formally by the following assumption.

\begin{assumption}
\label{multi_assum1_rct}
(Progressive process). The multistate processes \( Y_i^{(a)}(t) \) are non-decreasing in time, i.e., \( Y_i^{(a)}(t) \leq Y_i^{(a)}(s) \) for all \( 0 \leq t \leq s \), and satisfy \( Y_i^{(a)}(0) = 0 \).
\end{assumption}
Furthermore, we assume that the final state \( Q+1 \) is absorbing, meaning that once entered, the individual remains in that state permanently. %We also assume no competing risks interfere with this transition.

\begin{assumption}
\label{multi_assum2_rct}
(Single absorbing state). The state \( Q+1 \) is absorbing, and there are no competing risks that preclude entry into this state.
\end{assumption}

Under these assumptions, we first review the identification result developed in Mao.\cite{mao2023restricted} Specifically, we define the stage-specific transition times \( T_i^{q,(a)} = \inf\{t \geq 0 : Y_i^{(a)}(t) \geq q\}, \quad \text{for } q = 1,\dots,Q+1 \), which represent the earliest time at which the individual reaches state \( q \) or higher under treatment \( a \). Due to the progressive structure of the process, these transition times satisfy the ordering:
\( T_i^{1,(a)} \leq T_i^{2,(a)} \leq \dots \leq T_i^{Q+1,(a)}\). To compare the multistate outcomes between arms, one can consider the potential trajectories of individuals in each arm and assess how often one process dominates the other in terms of remaining in more favorable states. Specifically, we define the pairwise cumulative win time function, which quantifies the total duration over a fixed time horizon \( t \), during which one individual remains in a better state than another:
\[
\mathcal{W}\{Y_i^{(a)}, Y_j^{(1-a)}\}(t) = \int_{0}^{t} \mathbb{I}\{Y_i^{(a)}(u) < Y_j^{(1-a)}(u)\} \, du.
\]

The expected cumulative win time for the treatment group,  over all pairs across arms, is then defined as
\[
\xi^{(a)}(t) = \mathbb{E}\left[ \mathcal{W}\{Y_i^{(a)}, Y_j^{(1-a)}\}(t) \right],
\]
where the expectation is taken over independently sampled pairs of individuals, one from each potential outcome distribution. The causal estimand of interest, the restricted mean time in favor (RMT-IF) of treatment is then the difference in expected cumulative win times between arms:
\[
\Delta^{\text{rmt-if}}(t) = \xi^{(1)}(t) - \xi^{(0)}(t).
\]
To provide a more detailed examination of the treatment effect, Mao\cite{mao2023restricted} further decomposed the win time by state transitions. For each severity level \( q = 1,\dots,Q+1 \), we define the stage-specific win time function $\mathcal{W}^q\{Y_i^{(a)}, Y_j^{(1-a)}\}(t) = \int_{0}^{t} \mathbb{I}\{Y_i^{(a)}(u) < q, \, Y_j^{(1-a)}(u) = q\} \, du$, which quantifies the time individual \( i \) remains in a less severe state relative to individual \( j \)'s occupation of state \( q \). Summing across all stages, one then recovers the total win time as $\mathcal{W}\{Y_i^{(a)}, Y_j^{(1-a)}\}(t) = \sum_{q=1}^{Q+1} \mathcal{W}^q\{Y_i^{(a)}, Y_j^{(1-a)}\}(t)$. This leads to the stage-wise decomposition of the RMT-IF estimand:
\[
\Delta^{\text{rmt-if}}(t) = \sum_{q=1}^{Q+1} \Delta^{q,\text{rmt-if}}(t),
\]
where $\Delta^{q,\text{rmt-if}}(t) = \xi^{q,(1)}(t) - \xi^{q,(0)}(t)$ and $\xi^{q,(a)}(t) = \mathbb{E}\left[ \mathcal{W}^q\{Y_i^{(a)}, Y_j^{(1-a)}\}(t) \right]$. 

To obtain a a more operationalizable expression for the stage-wise estimand \(\Delta^{q,\text{rmt-if}}(t)\), one can define the expected win time \(\xi^{q,(a)}(t)\) in terms of the stage-specific transition times \(T_i^{q,(a)}\). For each severity level \(q = 1, \dots, Q+1\), define the marginal survival function \(S^{q,(a)}(t) = \mathbb{P}\left(T_i^{q,(a)} > t\right)\),
which denotes the probability that an individual in arm \(a\) remains in a state less severe than \(q\) at time \(t\). Similarly, the probability of being in exactly state \(q\) at time \(t\) is given by \(S^{q+1,(a)}(t) - S^{q,(a)}(t)\), under the progressive structure of the state process. Under randomization and independence between individuals across treatment arms, and under Assumptions \ref{multi_assum1_rct} and \ref{multi_assum2_rct}, the expected stage-wise win time can be written as
\[
\xi^{q,(a)}(t) = \int_0^t S^{q,(a)}(u) \left\{ S^{q+1,(1-a)}(u) - S^{q,(1-a)}(u) \right\} \, du.
\]
Built upon this, Mao\cite{mao2023restricted} developed the following expression for identification of RMT-IF:
\begin{proposition}
[Mao, 2023]\label{prop_stagewise_rct}
Under Assumptions \ref{multi_assum1_rct} and \ref{multi_assum2_rct}, the stage-wise RMT-IF estimand can be expressed as
\[
\Delta^{q,\text{rmt-if}}(t) = \int_0^t \left\{ S^{q,(1)}(u) S^{q+1,(0)}(u) - S^{q,(0)}(u) S^{q+1,(1)}(u) \right\} \, du,
\]
for all \(q = 1,\dots,Q+1\).
\end{proposition}

The representation of the stage-wise RMT-IF estimand captures the net expected duration over which treated individuals remain in more favorable states while control individuals occupy a specific severity level \(q\), relative to the reverse scenario. This formulation aggregates marginal survival probabilities across arms and time, and does not require modeling joint processes across individuals. %As such, it provides a transparent, stage-specific decomposition of treatment effect that aligns with the ordered progression of the multistate outcome. 
As a special case, when the state space contains only two levels, with \(Q = 1\) and state 1 denoting death, the transition time \(T_i^{1,(a)}\) corresponds to the survival time under treatment \(a\), and the stage-wise RMT-IF estimand reduces to the standard restricted mean survival time (RMST) contrast. That is, $\Delta^{\text{RMST}}(t) = \Delta^{\text{rmt-if}}(t) = \int_0^t \left\{ S^{(1)}(u) - S^{(0)}(u) \right\} du$, where $S^{(a)}(u) = \mathbb{P}(T_i^{(a)} > u)$. %Thus, the RMT-IF estimand extends the RMST framework to accommodate ordered multi-state outcome processes while preserving a similar marginal interpretation. \cite{mao2023restricted} %The proof is provided in Appendix XX.

\subsection{A covariate-adjusted doubly robust estimator for RMT-IF} \label{dr_estimator}

Estimation of the RMT-IF estimand \(\Delta^{\text{rmt-if}}(t)\), and its stage-wise components \(\Delta^{q,\text{rmt-if}}(t)\) in Proposition \ref{prop_stagewise_rct} requires estimation of the arm- and stage-specific survival functions \(S^{q,(a)}(t)\), for \(q=1,\dots, Q+1\), \(a \in \{0,1\}\). In practice, the transition times \(T_{i}^{q,(a)}\) are subject to right censoring due to administrative end of follow-up or loss to contact. Let \(C_i^{(a)}\) denote the potential censoring time for individual \(i\) under treatment assignment \(a \in \{0,1\}\). Under individual randomization, we assume that treatment assignment is independent of the potential multistate outcome processes conditional on baseline covariates, that is, \(A_i \;\perp\; \{Y_i^{(0)}(t),\, Y_i^{(1)}(t), C_i^{(0)}, C_i^{(1)}\} \,\big|\, \bm{Z}_i\). The treatment mechanism can be characterized by the arm-specific propensity score \(\pi^{(a)}(\bm{Z}_i) = \mathbb{P}(A_i = a \mid \bm{Z}_i)\) for \(a \in \{0,1\}\), which satisfies \(0 < \pi^{(a)}(\bm{Z}_i) < 1\) and is known by design (e.g., constant under simple randomization, or a known function of \(\bm{Z}_i\) under stratified or covariate-adaptive randomization). We adopt the Stable Unit Treatment Value Assumption (SUTVA), meaning that each participant's observed transition and censoring times coincide with their potential outcomes under their realized treatment. Thus, \(T_i^q = A_i T_i^{q,(1)} + (1 - A_i) T_i^{q,(0)}\) and \(C_i = A_i C_i^{(1)} + (1 - A_i) C_i^{(0)}\). For each state \(q\), we observe \(U_i^q = \min(T_i^q, C_i)\), and the event indicator is \(\delta_i^q = \mathbb{I}(T_i^q \leq C_i)\). To ensure identifiability of the stage-specific survival functions, we impose the following assumption for censoring.

\begin{assumption}{(Covariate-dependent censoring)}
\label{assum_3}
For each treatment arm \(a\in\{0,1\}\),  \(C_i^{(a)} \perp \{T_i^{1,(a)}, \dots, T_i^{Q+1,(a)}\} \mid \bm{Z}_i\) and for all \(t\), \(\mathbb{P}(C_i^{(a)}\geq t \mid \bm{Z}_i) > 0\). Under randomization and SUTVA, this is equivalent to \(C_i \;\perp\; \{T_i^{1}, \dots, T_i^{Q+1}\} \,\big|\, (A_i=a, \bm{Z}_i)\), that is, within each arm the observed censoring time is conditionally independent of the full collection of observed transition times given baseline covariates.
\end{assumption}

Let \(\bm{U}_i = \{U_i^1, \dots, U_i^{Q+1}\}\) and \(\bm{\delta}_i = \{\delta_i^1, \dots, \delta_i^{Q+1}\}\) denote the full collection of observed transition times and event indicators for participant \(i\). The observed data are \(O_i = \{A_i, \bm{Z}_i, Y_i, \bm{U}_i, \bm{\delta}_i\}\), which is independent and identically distributed across \(i=1,\dots, N\). The censoring mechanism in arm \(a\) is described as a conditional survival function \(K_c^{(a)}(t\mid \bm{Z}_i) = \mathbb{P}(C_i \geq t\mid A_i=a,\bm{Z}_i)\). By individual randomization and Assumption \ref{assum_3}, the probability that the potential transition time \(T_i^{q,(a)}\) is observed in arm \(a\) factorizes as \(P(A_i=a,C_i \geq T_i^{q,(a)}\mid \bm{Z}_i)  = \pi^{(a)}(\bm{Z}_i) K_c^{(a)}(T_{i}^{q,(a)}\mid \bm{Z}_i)\), which characterize the monotone coarsening mechanism for the stage \(q\)- transition time in arm \(a\). In this setting, \(S^{q,(a)}(t)\) admits an augmented inverse probability weighted (AIPWCC) representation. Building on the semiparametric theory of Robins and Rotnitzky \cite{robins1992recovery} and Tsiatis \cite{tsiatis2006semiparametric}, and adapting the influence function proposed by Bai et al. \cite{bai2013doubly} to the stage-specific multistate outcome, the estimating function for \(S^{q,(a)}(t)\) can be written as

\begin{align} \label{est_ee_irt}
    \phi(\bm{O}_{i})^{q,(a)}(t) &=  \frac{A_i^{a} (1-A_i)^{1-a} \delta_{i}^{q}}{\pi^{(a)}K_c^{(a)}(t\mid\bm{Z}_{i})} \mathbb{I}(U_{i}^{q}\geq t) 
    - \left\{\frac{A_i^{a} (1-A_i)^{1-a}  - \pi^{(a)}}{\pi^{(a)}} \right\} P(T_{i}^{q,(a)} \geq t|\bm{Z}_i) \nonumber \\
    & + \frac{A_i^{a} (1-A_i)^{1-a}}{\pi^{(a)}} \int_{0}^{\infty}  \frac{dM_c^{q,(a)}(u\mid\bm{Z}_{i})}{K_c^{(a)}(u\mid\bm{Z}_{i})} \frac{P(T_{i}^{q,(a)} \geq t|\bm{Z}_i )}{P(T_{i}^{q,(a)} \geq u|\bm{Z}_i )} - S^{q,(a)}(t) , 
\end{align} 
where \(M_c^{q,(a)}(u \mid \bm{Z}_{i}) = N_{i}^{q,C}(u) - \int_{0}^{u} \lambda_c^{(a)}(s \mid \bm{Z}_i) \mathbb{I}(U^{q} \geq s)\, ds\) is the martingale increment for the censoring process at stage \(q\), \( N_{i}^{q,C}(u) = \mathbb{I}(U_{i}^{q} \leq u, \delta_{i}^{q} = 0) \) denotes the counting process for censoring, and \(\lambda_c^{(a)}(u \mid \bm{Z}_i) = -{d \log K_c^{(a)}(u \mid \bm{z}_i)}/{du}\) is the hazard function for censoring given covariates. Here, the first term of \(\phi(O_i)^{q,(a)}(t)\) is the inverse probability weighted contribution of the observed survival indicator, scaled by the probability of receiving treatment \(a\) and remaining uncensored at time \(t\). The second and third terms together form the augmentation term. This augmentation is constructed so that the resulting influence function is orthogonal to the nuisance tangent space associated with the treatment and censoring mechanisms, and is motivated from the semiparametric efficient influence function for the marginal survival function for each transition time \(T^{q,(a)}\).\cite{westling2024inference} Thus, the corresponding estimator of \(S^{q,(a)}(t)\), denoted as \(\widehat{S}^{q,(a)}(t)\), is doubly robust. It is consistent if either (i) the censoring model \(K_c^{(a)}(t)\) (this model is common to all stages \(q=1,\dots,Q+1\)) or the outcome regression \(P(T_{i}^{q,(a)} \geq t|\bm{Z}_i )\) is correctly specified for \(a\in\{0,1\}\) and achieves the efficiency bound for this marginal survival function for the single stage \(q\). 

The estimator for \(S^{q,(a)}(t)\) can be obtained by solving the estimating equation \(\sum_{i=1}^{N} N^{-1} \phi(\bm{O}_i)^{^{q,(a)}}(t) = 0 \), which gives a closed-form representation
\begin{align} \label{stage_wise_S}
    \widehat{S}^{q,(a)}(t) &= \frac{1}{N}\sum_{i=1}^{N} \frac{A_i^{a} (1-A_i)^{1-a} \mathbb{I}(U_{i}^q \geq t) }{\pi^{(a)} \widehat{K}_c^{(a)}(t\mid\bm{Z}_{i})} - \left\{ \frac{A_i^{a} (1-A_i)^{1-a} - \pi^{(a)}}{ \pi^{(a)}} \right\}\widehat{P}(T_{i}^{q,(a)} \geq t\mid\bm{Z}_{i})   \nonumber\\
    & + \frac{A_i^{a} (1-A_i)^{1-a} }{\pi^{(a)}} \int_{0}^{t} \frac{d\widehat{M}_{c}^{q,(a)}(u\mid\bm{Z}_{i}) }{\widehat{K}_c^{(a)}(u\mid \bm{Z}_{i})} \frac{\widehat{P}(T_{i}^{q,(a)} \geq t|\bm{Z}_{i})}{\widehat{P}(T_{i}^{q,(a)}\geq u\mid\bm{Z}_{i} )}.
\end{align}
In (\ref{stage_wise_S}), \(\widehat{K}_c^{(a)}(t \mid \bm Z_i)\) denotes an estimator of the arm-specific censoring survival function \(K_c^{(a)}(t \mid \bm Z_i)\), and \(\widehat{P}(T_{i}^{q,(a)} \ge t \mid \bm Z_i)\) denotes an estimator of the survival function for the stage-\(q\) transition time under arm \(a\). Thus, \(\widehat{S}^{q,(a)}(t)\) is obtained by plugging in estimators of these two nuisance functions. A convenient default for modeling censoring is the arm-specific Cox proportional hazards model
\[
\lambda_c^{(a)}(t \mid \bm Z_i) = \lambda_{c0}^{(a)}(t)\exp(\bm{\gamma}_c^\top \bm Z_i),
\]
where \(\lambda_{c0}^{(a)}(t)\) is an unspecified baseline hazard and \(\bm{\gamma}_c\) is a vector of regression coefficients. The model is fit separately within each arm, and the Breslow estimator of \(\lambda_{c0}^{(a)}(t)\) yields \(\widehat{K}_c^{(a)}(t \mid \bm Z_i)\). For the stage-specific outcome regression \(P(T_{i}^{q,(a)} \ge t \mid \bm Z_i)\), an analogous Cox model may be specified for each stage \(q\),
\[
\lambda_q^{(a)}(t \mid \bm Z_i) = \lambda_{q0}^{(a)}(t)\exp(\bm{\beta}_q^\top \bm Z_i),
\]
where \(\lambda_{q0}^{(a)}(t)\) is the baseline hazard for the \(q\)th transition in arm \(a\) and \(\bm{\beta}_q\) is the associated regression parameter. The model is fit separately for each \(q\) and \(a\) using the right-censored data \((U_i^q,\delta_i^q)\) among individuals with \(A_i=a\), and the resulting fitted survival functions provide \(\widehat{P}(T_{i}^{q,(a)} \ge t \mid \bm Z_i)\). Other modeling choices for \(\widehat{K}_c^{(a)}\) and \(\widehat{P}\) may also be used. Semiparametric and parametric alternatives include accelerated failure time models and fully parametric survival models. In multistate settings, transition-specific Markov or semi-Markov models can be specified for each potential transition \(q\), with the resulting hazards combined using product integrals or the Aalen--Johansen estimator to obtain \(\widehat{S}^{q,(a)}(t)\) while preserving the ordering \(T_i^{1,(a)} \le \cdots \le T_i^{Q+1,(a)}\).\cite{cook2018multistate,aalen1978empirical} More specifically, let $\lambda_{rs}^{(a)}(t\mid \bm Z_i)$ denote the hazard of transitioning from state $r$ to state $s$ at time $t$ in arm $a$, for any transition under the progressive structure $(r<s)$. Given fitted transition-specific hazards $\widehat{\lambda}_{rs}^{(a)}(t\mid \bm Z_i)$, define the corresponding cumulative hazards $\widehat{\Lambda}_{rs}^{(a)}(t\mid \bm Z_i)=\int_0^t \widehat{\lambda}_{rs}^{(a)}(u\mid \bm Z_i)\,du$. These estimated cumulative hazards can then be combined over time to propagate probability mass across states starting from $Y_i^{(a)}(0)=0$, yielding state occupation probabilities $\widehat{p}_r^{(a)}(t\mid \bm Z_i)=\widehat{\mathbb{P}}(Y_i^{(a)}(t)=r\mid \bm Z_i)$ for $r=0,1,\dots,Q+1$ (equivalently, via the Aalen--Johansen construction in the nonparametric way). The stage-specific survival function is then estimated by
\[
\widehat{S}^{q,(a)}(t\mid \bm Z_i)=\sum_{r=0}^{q-1}\widehat{p}_r^{(a)}(t\mid \bm Z_i).
\]
This representation implies the nesting property $\widehat{S}^{q,(a)}(t\mid \bm Z_i)\ge \widehat{S}^{q+1,(a)}(t\mid \bm Z_i)$ for all $t$, because $\widehat{S}^{q+1,(a)}(t\mid \bm Z_i)=\widehat{S}^{q,(a)}(t\mid \bm Z_i)+\widehat{p}_q^{(a)}(t\mid \bm Z_i)$.
Since each stage-specific RMT-IF component in Proposition \ref{prop_stagewise_rct} is a smooth functional of the pair of marginal survival functions \(\{S^{q,(a)}(t), S^{q+1,(a)}(t)\}\), the plug-in estimator for \(\Delta^{q,\text{rmt-if}}(t)\) is then obtained by
\[
\widehat{\Delta}^{q,\text{rmt-if}}(t) = \int_{0}^{t} \left\{ \widehat{S}^{q,(1)}(u)\,\widehat{S}^{q+1,(0)}(u) - \widehat{S}^{q,(0)}(u)\,\widehat{S}^{q+1,(1)}(u)  \right\} \, du
\]
In particular, \(\widehat{\Delta}^{q,\text{rmt-if}}(t)\) is consistent if either all arm-specific censoring models are correctly specified or the stage-specific outcome regressions for \(T^{q,(a)}\) and \(T^{q+1,(a)}\) are correctly specified. In practice, the time integrals appearing in the stage-wise estimators can be computed numerically. We choose a grid of time points (e.g., observed time points) between \(0\) and the truncation time \(t\) as \(0 = t_0 < t_1 < \cdots < t_L = t\), and evaluate the estimated survival functions \(\widehat{S}^{q,(a)}(\cdot)\) at each grid point. The integral defining the stage-wise RMT-IF estimator can then be approximated by a finite sum over consecutive intervals using the trapezoidal rule:
\[
\widehat{\Delta}^{q,\text{rmt-if}}(t)
\approx \sum_{\ell=1}^{L}
\frac{
\widehat{S}^{q,(1)}(t_{\ell-1})\,\widehat{S}^{q+1,(0)}(t_{\ell-1})
- \widehat{S}^{q,(0)}(t_{\ell-1})\,\widehat{S}^{q+1,(1)}(t_{\ell-1})
+
\widehat{S}^{q,(1)}(t_{\ell})\,\widehat{S}^{q+1,(0)}(t_{\ell})
- \widehat{S}^{q,(0)}(t_{\ell})\,\widehat{S}^{q+1,(1)}(t_{\ell})
}{2}\,\{t_\ell - t_{\ell-1}\}.
\]
The overall RMT-IF estimator is obtained by summing over stages, \(\widehat{\Delta}^{\text{rmt-if}}(t) = \sum_{q=1}^{Q+1} \widehat{\Delta}^{q,\text{rmt-if}}(t)\). Since this is a linear combination of the stage-wise functionals, its doubly robust property follows directly from those of the underlying survival function estimators.
% Due to the form of this estimator, $\widehat{\Delta}^{q,\text{rmt-if}}(t)$ inherits the double robustness of the survival estimators under the same union model. 

\begin{proposition}
\label{prop:dr_irt_rmtif}
%Under randomized treatment assignment with known propensity scores, consider the plug-in estimator \(\widehat{\Delta}^{\text{rmt-if}}(t) = \sum_{q=1}^{Q+1} \widehat{\Delta}^{q,\text{rmt-if}}(t), \qquad t \ge 0\), constructed from the stage-specific survival estimators \(\widehat{S}^{q,(a)}(t)\) in \eqref{stage_wise_S}.   
Under individual randomization and covariate-dependent censoring, for any fixed \(t\), \(\widehat{\Delta}^{\text{rmt-if}}(t)\) is a consistent estimator of \(\Delta^{\text{rmt-if}}(t)\) if either (i) the common censoring model \(K_c^{(a)}(t \mid \bm Z_i)\) is correctly specified, or (ii) all stage-specific outcome model for \(P\{T_i^{q,(a)} \ge t \mid \bm Z_i\}\) are correctly specified for every \(q = 1,\dots,Q+1\) and \(a \in \{0,1\}\). 
\end{proposition}

Proof is provided in Web Appendix Section A. The double robustness result in Proposition \ref{prop:dr_irt_rmtif} further implies the following model robustness property under (arm-specific) independent censoring.

\begin{remark}
\label{rmk:model_robust_iid}
Under individual randomization and Assumption \ref{assum_3}, suppose the censoring process satisfies either of the following covriate-independent conditions:
\begin{enumerate}[(i)]
    \item \(C_i \;\perp\; \bm{Z}_i \mid A_i\), so that \(K_c^{(a)}(t \mid \bm Z_i) = K_c^{(a)}(t)\) and the arm-specific censoring distributions can be consistently estimated nonparametrically using arm-specific Kaplan--Meier estimators; or
    \item \(C_i \;\perp\; (A_i, \bm Z_i)\), so that \(K_c^{(a)}(t \mid \bm Z_i) = K_c(t)\) for all \(a\), and the common censoring distribution can be consistently estimated using a pooled Kaplan--Meier estimator.
\end{enumerate}
Then, for any choice of working models for the stage-specific outcome regressions \(P\{T_i^{q,(a)} \ge t \mid \bm Z_i\}\), \(q = 1,\dots,Q+1\), \(a \in \{0,1\}\), the plug-in estimator \(\widehat{\Delta}^{\text{rmt-if}}(t)\) remains a consistent estimator of \(\Delta^{\text{rmt-if}}(t)\). In particular, under independent censoring, the proposed estimator becomes model robust and the consistency to the RMT-IF estimand holds irrespective to the survival outcome model specification. 
\end{remark}

Under these independent censoring scenarios, the proposed estimator also provides a covariate-adjusted estimator compared to the unadjusted plug-in in Mao \cite{mao2023restricted}, and the following remark connects between these estimators. 
\begin{remark}
\label{rem:mao_connection}
Let \(\widehat{S}_{\mathrm{KM}}^{q,(a)}(t)\) denote the arm- and stage-specific Kaplan--Meier estimator of \(S^{q,(a)}(t)=\mathbb{P}\{T_i^{q,(a)}>t\}\) for \(q=1,\dots,Q+1\) and \(a\in\{0,1\}\). The estimator in Mao\cite{mao2023restricted} can be written as
\[ \widehat{\Delta}^{q,\mathrm{rmt\text{-}if}}_{\mathrm{Mao}}(t) = \int_{0}^{t} \!\Big\{ \widehat{S}_{\mathrm{KM}}^{q,(1)}(u)\,\widehat{S}_{\mathrm{KM}}^{q+1,(0)}(u) - \widehat{S}_{\mathrm{KM}}^{q,(0)}(u)\,\widehat{S}_{\mathrm{KM}}^{q+1,(1)}(u)
\Big\}\,du, \qquad \widehat{\Delta}^{\mathrm{rmt\text{-}if}}_{\mathrm{Mao}}(t) = \sum_{q=1}^{Q+1} \widehat{\Delta}^{q,\mathrm{rmt\text{-}if}}_{\mathrm{Mao}}(t).
\]
When baseline covariates are prognostic for the transition times \(T_i^{q,(a)}\), \(\widehat{\Delta}^{\mathrm{rmt\text{-}if}}(t)\) provides a more efficient estimator than \(\widehat{\Delta}^{\mathrm{rmt\text{-}if}}_{\mathrm{Mao}}(t)\) \cite{mao2023restricted}, and becomes asymptotically equivalent to \(\widehat{\Delta}^{\mathrm{rmt\text{-}if}}_{\mathrm{Mao}}(t)\) when covariates are uninformative of the multi-state outcome.
\end{remark}

\subsection{Variance estimation via group-jackknifing}
\label{jk_var_iid}

To estimate the variance of the doubly robust estimator of the restricted mean time in favor, \(\Delta^{\text{rmt-if}}(t)=\xi^{(1)}(t)-\xi^{(0)}(t)\), we consider a jackknife procedure directly to the pair \(\{\xi^{(1)}(t),\xi^{(0)}(t)\}\). The traditional leave-one-subject-out jackknife requires recomputing the estimator after removing each individual, leading to \(N\) replications and substantial computational burden. To simplify and accelerate computation, we instead adopt a group jackknife. That is, the \(N\) individuals are first partitioned into \(K\) approximately equal, non-overlapping subsets \(\{\mathcal{I}_1,\dots,\mathcal{I}_K\}\). Following Kott,\cite{kott2001delete} at least \(K = 15\) groups are recommended in practice. For each \(k\in\{1,\dots,K\}\), all individuals in \(\mathcal{I}_k\) are removed from the dataset, the nuisance functions \(\{K_c^{(a)}(t\mid\bm{Z}_i),\,P(T_i^{q,(a)}\ge t\mid\bm{Z}_i)\}\) are re-estimated on the remaining \(N-|\mathcal{I}_k|\) observations, and the arm-specific functionals \(\widehat{\xi}^{(a),-k}(t)\), \(a\in\{0,1\}\), are recomputed. This refitting procedure targets the variance of the full plug-in estimator, including the contribution from nuisance parameter estimation regardless of the model used, without requiring an explicit analytic derivation of the variance \cite{efron1981jackknife}. When  \(K=N\), and each  \(\mathcal{I}_k\) contains single individual, this procedure reduces to the leave-one-subject-out jackknife. Define the arm-specific jackknife means
\[
\bar{\xi}^{(a)}(t)=\frac{1}{K}\sum_{k=1}^{K}\widehat{\xi}^{(a),-k}(t),\qquad a\in\{0,1\},
\]
and the stacked vectors \(\widehat{\boldsymbol{\xi}}^{-k}(t)=\big(\widehat{\xi}^{(1),-k}(t),\,\widehat{\xi}^{(0),-k}(t)\big)^\top\) with jackknife mean  \(\bar{\boldsymbol{\xi}}(t)=\left\{\bar{\xi}^{(1)}(t),\,\bar{\xi}^{(0)}(t)\right\}^\top\). 
The jackknife covariance matrix for \(\{\xi^{(1)}(t),\xi^{(0)}(t)\}\) is
\[
\widehat{\boldsymbol{\Sigma}}_{\xi,\text{JK}}(t)
=\frac{K-1}{K}\sum_{k=1}^{K}\big\{\widehat{\boldsymbol{\xi}}^{-k}(t)-\bar{\boldsymbol{\xi}}(t)\big\}\big\{\widehat{\boldsymbol{\xi}}^{-k}(t)-\bar{\boldsymbol{\xi}}(t)\big\}^\top.
\]
The jackknife variance of \(\widehat{\Delta}^{\text{rmt-if}}(t)=\widehat{\xi}^{(1)}(t)-\widehat{\xi}^{(0)}(t)\) is obtained as the quadratic form
\begin{align} \label{var_iid}
    \widehat{\text{Var}}\big\{\widehat{\Delta}^{\text{rmt-if}}(t)\big\}
=\begin{pmatrix}1&-1\end{pmatrix}\widehat{\boldsymbol{\Sigma}}_{\xi,\text{JK}}(t)\begin{pmatrix}1\\-1\end{pmatrix}
\end{align}
Under standard regularity conditions ensuring asymptotic linearity, pointwise Wald confidence intervals follow from the normal distribution
\[
\widehat{\Delta}^{\text{rmt-if}}(t)\ \pm\ t_{1-\alpha/2,K-1}\,\sqrt{\widehat{\text{Var}}\big\{\widehat{\Delta}^{\text{rmt-if}}(t)\big\}},
\]
where \(t_{1-\alpha/2,K-1}\) is the \((1-\alpha/2)\)-quantile of the standard \(t\) distribution with \(K-1\) degree of freedom. %This approach avoids analytic variance derivations and remains valid irrespective of the specific working models used for the nuisance functions in the doubly robust estimator for RMT-IF. 
Because the jackknife is used only to estimate variance and does not modify the plug-in point estimate, the nominal coverage of the resulting confidence interval is driven by the doubly robust properties of the point estimator in Proposition~\ref{prop:dr_irt_rmtif}.

\section{Extension to cluster-randomized trials} \label{crt_estimand}

%In the previous section, we introduced the RMT-IF estimand and its doubly robust estimators for RCTs with multi-state survival outcomes. In RCTs, treatment assignment is independent at the individual level, so a single marginal estimand is sufficient to capture the treatment effect. 
In many practical settings, individual-level randomization is infeasible, for example, when interventions are delivered to entire groups such as hospitals, schools, or communities, or when there is a high risk of contamination between individuals. In such cases, CRTs are considered as a viable option.\cite{murray1998design,turner2017review1} While the potential outcome framework for RCTs extends naturally to CRT, the hierarchical data structure introduces additional complexity in defining the treatment effect. In particular, treatment effects can be defined under different weighting schemes: \cite{kahan2023estimands,kahan2023informative} the cluster-average treatment effect, which assigns equal weight to each cluster regardless of size, and the individual-average treatment effect, which assigns equal weight to each individual across all clusters. These estimands coincide in RCTs or when all clusters are of equal size, but may differ in the presence of informative cluster size, where cluster size is marginally associated with the cluster-specific treatment effect.\cite{li2025model} In what follows, we extend the RMT-IF framework to CRTs and propose the modified doubly robust methods for estimation. %by formalizing both the cluster-average RMT-IF and individual-average RMT-IF estimands.

\subsection{Formalizing the treatment effect estimands in CRTs with multi-state survival outcomes} 

Here, we use updated notation suitable for the CRT setting. We consider a CRT with \(M\) independent clusters indexed by \(i=1,\dots, M\), where cluster \(i\) contains \(N_i\) individuals indexed by \(j=1,\dots, N_i\). Each cluster is associated with a vector of cluster-level covariates \(\bm{W}_i\), and each individual \(j\) within cluster \(i\) has a vector of individual-level covariates \(\bm{Z}_{ij}\). We define \(\bm{V}_{ij} = \{\bm{W}_i, \bm{Z}_{ij}\}\) as the collection of all baseline covariates for participant \(j\) in cluster \(i\). Let \(A_i \in \{0,1\}\) denote the treatment assignment for cluster \(i\), with \(A_i=1\) indicating the active intervention and \(A_i=0\) indicating control. Extending the notation from the RCT setting, the potential multi-state outcome process for individual \(j\) in cluster \(i\) under treatment \(a\) is \(Y_{ij}^{(a)}(t) \in \{1,\dots, Q+1\}\), where larger states indicate worse status and \(Q+1\) denotes a terminal absorbing state such as death. Under the same assumptions as in \ref{multi_assum1_rct} and \ref{multi_assum2_rct}, the stage-specific potential transition time into state \(q\) or higher is  
\(T_{ij}^{q,(a)} = \inf \{ t \geq 0 : Y_{ij}^{(a)}(t) \geq q \}, \quad q=1,\dots,Q+1\),  
which satisfies \(T_{ij}^{1,(a)} \leq \dots \leq T_{ij}^{Q+1,(a)}\) by monotonicity. Following the RCT case, for an individual \(j\) in cluster \(i\) under arm \(a\) and an individual \(l\) in a distinct cluster \(k \neq i\) under arm \(1-a\), we define the pairwise cumulative win time as  
\[
\mathcal{W}\{ Y_{ij}^{(a)}, Y_{kl}^{(1-a)} \}(t) = \int_{0}^{t} \mathbb{I} \{Y_{ij}^{(a)}(u) < Y_{kl}^{(1-a)}(u) \} \, du.
\]
Aggregating these pairwise win times under different weighting schemes gives two natural CRT analogues of the RMT-IF. The cluster-level RMT-IF, which weights each cluster equally, is defined as
\[
\xi_C^{(a)}(t) = \mathbb{E} \left[ \frac{\sum_{j=1}^{N_i} \sum_{l=1}^{N_k} \mathcal{W}\{ Y_{ij}^{(a)}, Y_{kl}^{(1-a)} \}(t)}{N_i N_k} \right], 
\quad \Delta_C^{\text{rmt-if}}(t) = \xi_C^{(1)}(t) - \xi_C^{(0)}(t).
\]
The individual-level RMT-IF, which weights each individual equally across clusters, is defined as
\[
\xi_I^{(a)}(t) = \frac{\mathbb{E} \left[ \sum_{j=1}^{N_i} \sum_{l=1}^{N_k} \mathcal{W}\{ Y_{ij}^{(a)}, Y_{kl}^{(1-a)} \}(t) \right]}{\mathbb{E}(N_i N_k)}, 
\quad \Delta_I^{\text{rmt-if}}(t) = \xi_I^{(1)}(t) - \xi_I^{(0)}(t).
\]
These estimands generalize the average treatment effect estimands defined in Li et al.\cite{li2025model} to pairwise comparison regimes. When all clusters are of equal size, or when cluster size is non-informative, these two estimands coincide, \(\Delta_C^{\text{rmt-if}}(t) = \Delta_I^{\text{rmt-if}}(t)\), and reduce to the single RMT-IF in the RCT case. In the special case \(Q=1\), they collapse to the corresponding cluster- and individual-level RMST contrasts. Similarly to the stage-wise decomposition, we can further decompose the win time by state transitions as 
\[
\mathcal{W}\{Y_{ij}^{(a)}, Y_{kl}^{(1-a)}\}(t) = \sum_{q=1}^{Q+1} \mathcal{W}^q\{Y_{ij}^{(a)}, Y_{kl}^{(1-a)}\}(t).
\]
This leads to the stage-wise decomposition of the CRT estimands as $\Delta_C^{\text{rmt-if}}(t) = \sum_{q=1}^{Q+1} \Delta_C^{q,\text{rmt-if}}(t)$ and $\Delta_I^{\text{rmt-if}}(t) = \sum_{q=1}^{Q+1} \Delta_I^{q,\text{rmt-if}}(t)$, 
% \[
% \Delta_C^{\text{rmt-if}}(t) = \sum_{q=1}^{Q+1} \Delta_C^{q,\text{rmt-if}}(t), \quad 
% \Delta_I^{\text{rmt-if}}(t) = \sum_{q=1}^{Q+1} \Delta_I^{q,\text{rmt-if}}(t),
% \]
where \( \Delta_C^{q,\text{rmt-if}}(t) = \xi_C^{q,(1)}(t) - \xi_C^{q,(0)}(t) \), \( \Delta_I^{q,\text{rmt-if}}(t) = \xi_I^{q,(1)}(t) - \xi_I^{q,(0)}(t)\), and
\[
\xi_C^{q,(a)}(t) = \mathbb{E} \left[ \frac{\sum_{j=1}^{N_i} \sum_{l=1}^{N_k} \mathcal{W}^q\{Y_{ij}^{(a)}, Y_{kl}^{(1-a)}\}(t)}{N_i N_k} \right], \quad
\xi_I^{q,(a)}(t) = \frac{\mathbb{E} \left[ \sum_{j=1}^{N_i} \sum_{l=1}^{N_k} \mathcal{W}^q\{Y_{ij}^{(a)}, Y_{kl}^{(1-a)}\}(t) \right]}{\mathbb{E}(N_i N_k)}.
\]
To connect these with survival functions, we follow Fang et al.\cite{fang2025estimands} and define the cluster-level and individual-level stage-specific survival functions as
\[
S_C^{q,(a)}(t) = \mathbb{E} \left\{ \frac{1}{N_i} \sum_{j=1}^{N_i} \mathbb{I}\left(T_{ij}^{q,(a)} \ge t \right) \right\}, 
\quad S_I^{q,(a)}(t) = \frac{\mathbb{E} \left\{ \sum_{j=1}^{N_i} \mathbb{I}\left(T_{ij}^{q,(a)} \ge t \right) \right\}}{\mathbb{E}(N_i)}.
\]
Here, \(S_C^{q,(a)}(t)\) corresponds to the probability, averaged equally across clusters, that an individual in arm \(a\) has not yet reached state \(q\) by time \(t\), whereas \(S_I^{q,(a)}(t)\) corresponds to the same probability when weighting individuals equally across the full study population. For identification under cluster randomization, we prove the following results in Web Appendix B.

\begin{proposition}
\label{prop_stagewise_crt}
Under Assumptions \ref{multi_assum1_rct} and \ref{multi_assum2_rct}, the stage-wise CRT estimands have the representation
\[
\Delta_C^{q,\text{rmt-if}}(t) = \int_0^t \left\{ S_C^{q,(1)}(u) S_C^{q+1,(0)}(u) - S_C^{q,(0)}(u) S_C^{q+1,(1)}(u) \right\} \, du,
\]
\[
\Delta_I^{q,\text{rmt-if}}(t) = \int_0^t \left\{ S_I^{q,(1)}(u) S_I^{q+1,(0)}(u) - S_I^{q,(0)}(u) S_I^{q+1,(1)}(u) \right\} \, du,
\]
for all \(q = 1,\dots,Q+1\).
\end{proposition}

\subsection{Doubly robust estimators for analyzing RMT-IF estimands in CRT}

Estimation of the CRT estimands \(\Delta_C^{\text{rmt-if}}(t)\) and \(\Delta_I^{\text{rmt-if}}(t)\) is built upon their stage-wise representation in Proposition \ref{prop_stagewise_crt}, where the treatment effect is calculated from the marginal stage-specific survival functions \(S_C^{q,(a)}(t)\) and \(S_I^{q,(a)}(t)\) for \(q = 1,\dots,Q+1\). In practice, the transition times \(T_{ij}^{q,(a)}\) may be subject to right-censoring. Let \(C_{ij}^{(a)}\) denote the potential censoring time for individual \(j\) in cluster \(i\) under arm \(a\).  Under cluster randomization, the treatment assignment \(A_i\) is independent of all potential multistate outcomes and censoring processes conditional on baseline covariates, that is, \(A_i \perp \{Y_{ij}^{(0)}(t), Y_{ij}^{(1)}(t), C_{ij}^{(0)}, C_{ij}^{(1)} : j = 1,\dots, N_i\} \mid \bm{V}_i\), with arm-specific propensity scores \(0 < \pi^{(a)}(\bm{V}_i) = \mathbb{P}(A_i = a \mid \bm{V}_i) < 1\) for \(a \in \{0,1\}\), where \(\bm{V}_i = \{\bm{V}_{i1}, \dots, \bm{V}_{iN_i}\}\) is the collection of cluster-level and individual-level baseline covariates. We impose a cluster-level SUTVA, under which each individual's potential multistate and censoring processes depend only on the treatment assignment of their own cluster and the observed processes coincide with the corresponding processes under the realized treatment, thus, \(T_{ij}^{q} = A_i T_{ij}^{q,(1)} + (1 - A_i) T_{ij}^{q,(0)}\) for \(q = 1,\dots,Q+1\), and \(C_{ij} = A_i C_{ij}^{(1)} + (1 - A_i) C_{ij}^{(0)}\). We assume the following conditional independence for censoring at the cluster level.
\begin{assumption}{(Covariate-dependent censoring)}\label{assum_crt_cens}
For \(a \in \{0,1\}\), \(\{C_{i1}^{(a)}, \dots, C_{iN_i}^{(a)}\} \perp \{T_{i1}^{q,(a)}, \dots, T_{iN_i}^{q,(a)} : q = 1,\dots,Q+1\} \mid (\bm{V}_i, N_i)\). Under cluster randomization and cluster-level SUTVA, this is equivalent to \(\{C_{i1}, \dots, C_{iN_i}\} \perp \{T_{i1}^{q}, \dots, T_{iN_i}^{q} : q = 1,\dots,Q+1\} \mid (\bm{V}_i, N_i, A_i = a)\) within each arm \(a\).
\end{assumption}

Assumption \ref{assum_crt_cens} is formulated at cluster level, that is, conditional on \((\bm{V}_i, N_i)\), the joint censoring process \(\{C_{ij}^{(a)}\}_{j=1}^{N_i}\) is independent of the joint collection of stage-specific event times \(\{T_{ij}^{q,(a)}\}_{j=1, q=1}^{N_i, Q+1}\), but allows for arbitrary within-cluster correlation in both the multistate outcome and censoring processes. The observed time for state \(q\) is \(U_{ij}^q = \min(T_{ij}^q, C_{ij})\) with event indicator \(\delta_{ij}^q = \mathbb{I}(T_{ij}^q \le C_{ij})\). The full observed data for individual \(j\) in cluster \(i\) is \(O_{ij} = (A_i, \bm{W}_i, N_i, \bm{Z}_{ij}, \bm{U}_{ij}, \bm{\delta}_{ij})\), where \(\bm{U}_{ij} = \{U_{ij}^1, \dots, U_{ij}^{Q+1}\}\) and \(\bm{\delta}_{ij} = \{\delta_{ij}^1, \dots, \delta_{ij}^{Q+1}\}\). Let \(\bm{O}_i = \{O_{i1}, \dots, O_{iN_i}\}\) denote the collection of observations in cluster \(i\). Under the cluster randomization,  \(\bm{O}_1, \dots, \bm{O}_M\) are identically independently distributed across \(i = 1,\dots,M\). In the CRT setting, each cluster is treated as an independent data unit, and we adapt the AIPWCC estimator developed in Fang et al.\cite{fang2025estimands} to estimate the stage-specific survival function. The contribution of cluster \(i\) to the estimating equation for \(S_C^{q,(a)}(t)\) can be expressed as
\begin{align}
\phi_C(\bm{O}_i)^{q,(a)}(t) &= \frac{1}{N_i}\sum_{j=1}^{N_i} \Bigg[ \frac{A_i^{a}(1-A_i)^{1-a} \, \delta_{ij}^q}{\pi^{(a)}K_c^{(a)}(t \mid \bm{V}_i)} \, \mathbb{I}\{U_{ij}^q \ge t\} \nonumber \\
&\quad - \left\{\frac{A_i^{a}(1-A_i)^{1-a} - \pi^{(a)}}{\pi^{(a)}}\right\} \, P\left(T_{ij}^{q,(a)} \ge t \mid \bm{V}_i\right) \nonumber \\
&\quad + \frac{A_i^{a}(1-A_i)^{1-a}}{\pi^{(a)}} \int_0^\infty \frac{dM_c^{q,(a)}(u \mid \bm{V}_{ij})}{K_c^{(a)}(u \mid \bm{V}_i)} \, \frac{P\left(T_{ij}^{q,(a)} \ge t \mid \bm{V}_i\right)}{P\left(T_{ij}^{q,(a)} \ge u \mid \bm{V}_i\right)} \Bigg],
\label{AIPWCC_cluster_stagewise}
\end{align}
where \(M_c^{q,(a)}(u \mid \bm{O}_{ij}) = N_{ij}^{q,C}(u) - \int_0^u \lambda_c^{(a)}(s \mid \bm{V}_i) \mathbb{I}(U_{ij}^q \ge s)\, ds\) is the martingale increment for the censoring process, \(N_{ij}^{q,C}(u) = \mathbb{I}(U_{ij}^q \le u, \Delta_{ij}^q = 0)\) is the censoring counting process. The function \(K_c^{(a)}(t \mid \bm{V}_i)\) is the conditional survival function for censoring, and \(\lambda_c^{(a)}\) is its hazard. This form parallels the influence function in \eqref{est_ee_irt}, but averages over individuals within each cluster to target the cluster-level survival function. Under the working independence assumption for individuals within a cluster, the cluster-level survival function \(\widehat{S}_C^{q,(a)}(t)\) is obtained by solving the estimating equation \(\sum_{i=1}^M \phi_C(\bm{O}_i)^{q,(a)}(t) = 0\), with the closed-form
\begin{align*}
    \widehat{S}_C^{q,(a)}(t) &= \frac{1}{M} \sum_{i=1}^M \frac{1}{N_i} \sum_{j=1}^{N_i} \Bigg[ \frac{A_i^{a}(1-A_i)^{1-a} \, \mathbb{I}(U_{ij}^q \ge t)}{\pi^{(a)}\widehat{K}_c^{(a)}(t \mid \bm{V}_i)}
- \left\{ \frac{A_i^{a}(1-A_i)^{1-a} - \pi^{(a)}}{\pi^{(a)}} \right\} \widehat{P}\{T_{ij}^{q,(a)} \ge t \mid \bm{V}_i\} \\
& + \frac{A_i^{a}(1-A_i)^{1-a}}{\pi^{(a)}} \int_0^{t} \frac{d\widehat{M}_c^{q,(a)}(u \mid \bm{V}_{ij})}{\widehat{K}_c^{(a)}(u \mid \bm{V}_i)} \frac{\widehat{P}\{T_{ij}^{q,(a)} \ge t \mid \bm{V}_i\}}{\widehat{P}\{T_{ij}^{q,(a)} \ge u \mid \bm{V}_i\}} \Bigg].
\end{align*}
The estimator for individual-level estimators can be derived from \(\sum_{i=1}^{M}  \left\{N_i \phi_C(O_{ij})^{q,(a)}(t) + S_C^{q,(a)}(t) - S_I^{q,(a)}(t)  \right\} = 0 \) with the closed form
\[
\widehat{S}_I^{q,(a)}(t) = \frac{\sum_{i=1}^M N_i \, \widehat{S}_C^{q,(a)}(t)}{\sum_{i=1}^M N_i}.
\]

In the expressions above, \(\widehat{K}_c^{(a)}(t \mid \bm{V}_i)\) denotes an estimator of the arm-specific censoring survival function \(K_c^{(a)}(t \mid \bm{V}_i)\), and \(\widehat{P}\{T_{ij}^{q,(a)} \ge t \mid \bm{V}_i\}\) denotes an estimator of the survival function for the stage-\(q\) transition time under arm \(a\). For the censoring model, a convenient choice is an arm-specific marginal Cox proportional hazards model
\[
\lambda_{c,ij}^{(a)}(t \mid \bm{V}_i) = \lambda_{c0}^{(a)}(t)\,\exp\{\bm{\gamma}_c^\top \bm{V}_i\},
\]
where \(\lambda_{c0}^{(a)}(t)\) is an unspecified baseline hazard in arm \(a\), and \(\bm{\gamma}_c\) is a vector of regression coefficients. Alternatively, we can also choose a shared frailty Cox model to allow more flexible within-cluster dependence in the censoring process, 
\[
\lambda_{c,ij}^{(a)}(t \mid \bm{V}_i, B_i) = B_i\,\lambda_{c0}^{(a)}(t)\,\exp\{\bm{\gamma}_c^\top \bm{V}_i\},
\]
where \(B_i\) is a nonnegative cluster-level frailty term. In this case, \(K_c^{(a)}(t \mid \bm{V}_i)\) is obtained by integrating out the frailty distribution. The stage-specific outcome regressions \(P\{T_{ij}^{q,(a)} \ge t \mid \bm{V}_i\}\) can be modeled analogously. A marginal Cox model for the stage-\(q\) transition time in arm \(a\) takes the form
\[
\lambda_{q,ij}^{(a)}(t \mid \bm{V}_i) = \lambda_{q0}^{(a)}(t)\,\exp\{\bm{\beta}_q^\top \bm{V}_i\},
\]
where \(\lambda_{q0}^{(a)}(t)\) is the arm- and stage-specific baseline hazard and \(\bm{\beta}_q\) is the corresponding regression parameter. To incorporate additional within-cluster heterogeneity in the event process, we may specify shared frailty Cox models,
\[
\lambda_{q,ij}^{(a)}(t \mid \bm{V}_i, U_i^{(q)}) = R_i^{(q)}\,\lambda_{q0}^{(a)}(t)\,\exp\{\bm{\beta}_q^\top \bm{V}_i\},
\]
with cluster-level frailties \(R_i^{(q)}\), which is stage-specific. Additional details of these survival models along with the computational steps for constructing the doubly robust estimator are provided in Fang et al. for a single-state survival outcome.\cite{fang2025estimands} The same steps can be used for our estimator in the context of stage-specific survival function estimator. 

Both estimators \(\widehat{S}_C^{q,(a)}(t)\) and \(\widehat{S}_I^{q,(a)}(t)\) retain the doubly robust feature of the underlying AIPWCC formulation, and this property carries over to the CRT RMT-IF estimators obtained by plugging \(\widehat{S}_C^{q,(a)}(t)\) and \(\widehat{S}_I^{q,(a)}(t)\) into the stage-wise representation in Proposition \ref{prop_stagewise_crt}.

\begin{proposition}
\label{prop:dr_crt_rmtif}
Under cluster randomization, for any fixed \(t\), the plug-in estimators \(\widehat{\Delta}_C^{\text{rmt-if}}(t)\) and \(\widehat{\Delta}_I^{\text{rmt-if}}(t)\) based on \(\widehat{S}_C^{q,(a)}(t)\) and \(\widehat{S}_I^{q,(a)}(t)\) are consistent for \(\Delta_C^{\text{rmt-if}}(t)\) and \(\Delta_I^{\text{rmt-if}}(t)\) if either (i) for each arm \(a\), the censoring model \(K_c^{(a)}(t \mid \bm{V}_i)\) is correctly specified, or (ii) for all \(q = 1,\dots,Q+1\), the stage-specific outcome regressions \(P\{T_{ij}^{q,(a)} \ge t \mid \bm{V}_i\}\) are correctly specified.
\end{proposition}

In the special case, where the censoring is completely independent or arm-specific independent, we have the following model robust properties.
\begin{remark} \label{rmk:model_robust_crt}
Under cluster randomization and Assumption \ref{assum_crt_cens}, suppose the censoring process satisfies either of the following:
\begin{enumerate}[i]
    \item  \(\{C_{i1},\dots, C_{iN_i} \} \perp (\bm V_i, N_i) \mid A_i \), so that \(K_c^{(a)}(t \mid \bm{V}_i, N_i) = K_c^{(a)}(t)\), and the arm-specific censoring distributions \(K_c^{(a)}\) can be consistently estimated nonparametrically using arm-specific Kaplan--Meier estimators; or
    \item \(\{C_{i1},\dots, C_{iN_i} \} \perp  (A_i, \bm V_i,N_i) \), so that \(K_c^{(a)}(t \mid \bm{V}_i, N_i) = K_c(t)\) for all \(a\), and a common censoring distribution \(K_c\) can be consistently estimated using a pooled Kaplan--Meier estimator.
\end{enumerate}
Then, for any choice of models for the stage-specific outcome regressions  \(P\{T_{ij}^{q,(a)} \ge t \mid \bm{V}_i\}, \quad q = 1,\dots,Q+1, \quad a \in \{0,1\} \) the estimators \(\widehat{\Delta}_C^{\mathrm{rmt-if}}(t)\) and \(\widehat{\Delta}_I^{\mathrm{rmt-if}}(t)\) are consistent estimators of \(\Delta_C^{\mathrm{rmt-if}}(t)\) and \(\Delta_I^{\mathrm{rmt-if}}(t)\). In particular, under independent censoring, the proposed CRT RMT-IF estimators are model robust and their consistency holds irrespective of the specification of the multistate survival outcome models.
\end{remark}
Remark \ref{rmk:model_robust_crt} provides a natural generalization of the model-robust standardization framework for CRTs proposed by Li et al.\ \cite{li2025model}, extending it from non-survival outcomes (e.g., continuous, binary, and count) to multistate survival outcomes subject to right censoring.
%As in the IRT scenario, Proposition \ref{prop:dr_crt_rmtif} implies a model robustness property under (arm-specific) independent censoring, in which the censoring distribution can be estimated nonparametrically without specifying a regression model for \(K_c^{(a)}(t)\). 
In addition, when the prognostic covariates are non-informative for the multistate survival outcomes, this framework provides an extension of estimators in Mao \cite{mao2023restricted} to CRT. 
\begin{remark}\label{rmk:crt_km}
Let \(\widehat{S}_{I,KM}^{q,(a)}(t)\) denote the arm- and stage-specific Kaplan--Meier estimator of  \(S_I^{q,(a)}(t) = \mathbb{P}\{T_{ij}^{q,(a)} > t\},\quad q = 1,\dots,Q+1,\; a \in \{0,1\}\),
obtained by pooling all individuals within arm \(a\) and treating them as independent. A direct CRT analogue of the estimator in Mao \cite{mao2023restricted} based on individual-level data can be written as
\[
\widehat{\Delta}_{I,q,\mathrm{rmt\text{-}if}}^{KM}(t)  = \int_0^t \Big\{\widehat{S}_{I,KM}^{q,(1)}(u)\,\widehat{S}_{I,KM}^{q+1,(0)}(u)  - \widehat{S}_{I,KM}^{q,(0)}(u)\,\widehat{S}_{I,KM}^{q+1,(1)}(u) \Big\}\,du,\qquad \widehat{\Delta}_{I,\mathrm{rmt\text{-}if}}^{KM}(t)  = \sum_{q=1}^{Q+1} \widehat{\Delta}_{I,q,\mathrm{rmt\text{-}if}}^{KM}(t).\]
Similary a cluster-level Kaplan--Meier estimator \(\widehat{S}_{C,KM}^{q,(a)}(t)\) for  \(S_C^{q,(a)}(t)\) is obtained by fitting a weighted Kaplan--Meier within arm \(a\), assigning weight \(1/N_i\) to each individual in cluster \(i\) so that each cluster contributes total weight one. The corresponding cluster-level estimators are
\[
\widehat{\Delta}_{C,q,\mathrm{rmt\text{-}if}}^{KM}(t)  = \int_0^t \Big\{\widehat{S}_{C,KM}^{q,(1)}(u)\,\widehat{S}_{C,KM}^{q+1,(0)}(u)  - \widehat{S}_{C,KM}^{q,(0)}(u)\,\widehat{S}_{C,KM}^{q+1,(1)}(u) \Big\}\,du,\qquad \widehat{\Delta}_{C,\mathrm{rmt\text{-}if}}^{KM}(t)  = \sum_{q=1}^{Q+1} \widehat{\Delta}_{C,q,\mathrm{rmt\text{-}if}}^{KM}(t).
\]
\end{remark}
Thus, in CRTs, our RMT-IF estimators target the same marginal arm-specific multistate survival contrasts as these Kaplan--Meier plug-in analogues, but, when prognostic cluster-level or individual-level covariates are informative, they can leverage flexible outcome and censoring models to improve efficiency while retaining the desired model-robustness properties.

%The role of weighting is more delicate in CRTs than in IRTs when the prognostic covariates are noninformative. An arm-specific Kaplan--Meier estimator obtained by pooling all individuals within arm \(a\) and treating them as independent targets the individual-level survival function \(S_I^{q,(a)}(t)\), and, when substituted into the RMT-IF functional, gives an estimator for the individual-level RMT-IF \(\Delta_I^{\text{rmt-if}}(t)\). In contrast, the cluster-level survival function \(S_C^{q,(a)}(t)\) and the corresponding cluster-level RMT-IF \(\Delta_C^{\text{rmt-if}}(t)\) require equal weighting of clusters and need to apply weights \(1/N_i\) for each individual. As a result, a direct Kaplan--Meier plug-in estimator in Mao \cite{mao2023restricted} is not, in general, compatible with the cluster-average CRT estimand without appropriate reweighting. 

\subsection{Estimating the variance via cluster jackknifing} 
\label{jk_var_crt}
To estimate the variance of the doubly robust estimators \(\Delta_C^{\text{rmt-if}}(t)\) and \(\Delta_I^{\text{rmt-if}}(t)\) in the CRT setting, we apply a leave-one-cluster-out jackknife directly to the arm-specific functionals \(\{\xi_C^{(1)}(t),\xi_C^{(0)}(t)\}\) and \(\{\xi_I^{(1)}(t),\xi_I^{(0)}(t)\}\). For \(g \in \left\{1,\dots,M\right\}\), let \(\mathcal{C}_g\) be the set of all individuals in cluster \(g\). At each iteration \(g\), all observations in \(\mathcal{C}_g\) are removed, the nuisance functions \(\left\{K_c^{(a)}(t\mid\bm{V}_i),\,P\left(T_{ij}^{q,(a)} \ge t\mid\bm{V}_i\right)\right\}\) are re-estimated using the remaining \(M-1\) clusters, and the arm-specific functionals \(\widehat{\xi}_C^{(a),-g}(t)\) and \(\widehat{\xi}_I^{(a),-g}(t)\) for \(a \in \left\{0,1\right\}\) are recomputed. The jackknife means are \(\bar{\xi}_C^{(a)}(t)=M^{-1}\sum_{g=1}^M \widehat{\xi}_C^{(a),-g}(t)\) and \(\bar{\xi}_I^{(a)}(t)=M^{-1}\sum_{g=1}^M \widehat{\xi}_I^{(a),-g}(t)\). Define \(\widehat{\boldsymbol{\xi}}_C^{-g}(t)=\left(\widehat{\xi}_C^{(1),-g}(t),\,\widehat{\xi}_C^{(0),-g}(t)\right)^\top\) and \(\widehat{\boldsymbol{\xi}}_I^{-g}(t)=\left(\widehat{\xi}_I^{(1),-g}(t),\,\widehat{\xi}_I^{(0),-g}(t)\right)^\top\) with corresponding means \(\bar{\boldsymbol{\xi}}_C(t)=\left(\bar{\xi}_C^{(1)}(t),\,\bar{\xi}_C^{(0)}(t)\right)^\top\) and \(\bar{\boldsymbol{\xi}}_I(t)=\left(\bar{\xi}_I^{(1)}(t),\,\bar{\xi}_I^{(0)}(t)\right)^\top\).  

The jackknife covariance matrices are  
\[
\widehat{\boldsymbol{\Sigma}}_{C,\text{JK}}(t)=\frac{M-1}{M}\sum_{g=1}^M\left[\widehat{\boldsymbol{\xi}}_C^{-g}(t)-\bar{\boldsymbol{\xi}}_C(t)\right]\left[\widehat{\boldsymbol{\xi}}_C^{-g}(t)-\bar{\boldsymbol{\xi}}_C(t)\right]^\top,
\]
\[
\widehat{\boldsymbol{\Sigma}}_{I,\text{JK}}(t)=\frac{M-1}{M}\sum_{g=1}^M\left[\widehat{\boldsymbol{\xi}}_I^{-g}(t)-\bar{\boldsymbol{\xi}}_I(t)\right]\left[\widehat{\boldsymbol{\xi}}_I^{-g}(t)-\bar{\boldsymbol{\xi}}_I(t)\right]^\top.
\]

The jackknife variances of \(\widehat{\Delta}_C^{\text{rmt-if}}(t)=\widehat{\xi}_C^{(1)}(t)-\widehat{\xi}_C^{(0)}(t)\) and \(\widehat{\Delta}_I^{\text{rmt-if}}(t)=\widehat{\xi}_I^{(1)}(t)-\widehat{\xi}_I^{(0)}(t)\) are  
\[
\widehat{\text{Var}}\left\{\widehat{\Delta}_C^{\text{rmt-if}}(t)\right\}=\left(1,-1\right)\widehat{\boldsymbol{\Sigma}}_{C,\text{JK}}(t)\left(1,-1\right)^\top,
\]
\[
\widehat{\text{Var}}\left\{\widehat{\Delta}_I^{\text{rmt-if}}(t)\right\}=\left(1,-1\right)\widehat{\boldsymbol{\Sigma}}_{I,\text{JK}}(t)\left(1,-1\right)^\top.
\]
The leave-one-cluster-out approach preserves the correlation structure induced by cluster randomization, ensuring valid cluster-level inference without analytic variance derivations. As in the i.i.d. case, validity does not depend on the correct specification of all working models for the nuisance functions. To mitigate small-sample bias, Wald confidence intervals are constructed using the \(t\)-distribution with \(M-2\) degrees of freedom.\citep{li2017evaluation}

\section{Simulation studies} \label{sec:sim}

\subsection{Simulations under individual randomization}
We consider a two-arm IRT with a binary treatment indicator \(A_i\), where \(A_i=1\) if participant \(i\) is randomized to the experimental intervention and \(A_i=0\) otherwise, and \(\mathbb{P}(A_i=1)=\mathbb{P}(A_i=0)=0.5\). The total sample size is set to \(N=2000\). We generate two baseline covariates, \(Z_{i1}\) and \(Z_{i2}\), which are independently distributed as \(Z_{i1}\sim \text{Normal}(0,1)\) and \(Z_{i2}\sim \text{Bernoulli}(0.5)\). We consider a three-state progressive process as a validation and illustrative setting; the proposed method is not restricted to three states, where the third state is absorbing state (terminal event). %Event times are generated from Cox-type transition intensities model. 
The potential transition time from state 1, denoted \(\widetilde{T}_i^{1,(a)}\) under treatment assignment \(a\), is generated based on hazard function \(
\lambda_1^{(a)}(t\mid Z_{i1},Z_{i2}) = (0.2+0.2(1-a) )\exp\left\{-a + Z_{i1} +0.5Z_{i2} + Z_{i1}Z_{i2} \right\}.
\)
The transition from state 2 is generated based on a gap time \(u_i\), so that \(\widetilde{T}_i^{2,(a)}=\widetilde{T}_i^{1,(a)}+u_i\), with hazard \(
\lambda_u^{(a)}(u\mid Z_{i1},Z_{i2}) = \{0.5+0.5(1-a)\} \exp\left\{-1.5a+Z_{i1} + 0.5Z_{i2} +0.5 Z_{i1}Z_{i2} \right\}.
\)
Finally, the transition to the absorbing state (state 3), \(T_i^{3,(a)}\) is generated from \(
\lambda_3(t\mid Z_{i1},Z_{i2}) = \{0.1+0.05(1-a)\} \exp\left\{-a +0.5Z_{i1}+Z_{i2} + Z_{i1}Z_{i2}\right\}.
\) Censoring times \(C_i^{(a)}\) are generated independently from a hazard model \(
h(t\mid Z_{i1},Z_{i2})= 0.26 \exp\left\{2a -1.5Z_{i1} - Z_{i2} -2 Z_{i1}Z_{i2} \right\}.
\)
Event times are defined as \(T_i^{1,(a)}=\min\{\widetilde{T}_i^{1,(a)},\widetilde{T}_i^{2,(a)},T_i^{3,(a)}\}\) and \(T_i^{2,(a)}=\min\{\widetilde{T}_i^{2,(a)},T_i^{3,(a)}\}\). The observed time is \(U_i^{q,(a)}=T_i^{q,(a)}\wedge C_i^{(a)}\) for \(q\in\{1,2,3\}\), with status indicator \(\delta_i=1\) if \(T_i^{1,(a)}\leq C_i^{(a)}\), \(\delta_i=2\) if \(T_i^{2,(a)}\leq C_i^{(a)}\), and \(\delta_i=3\) if \(T_i^{3,(a)}\leq C_i\). The censoring rate is 50\% in this setting.

Our target estimands include \(\xi^{(a)}(t)\) for \(a=0,1\) and the RMT-IF \(\Delta^{\text{rmt-if}}(t)\) on the difference scale. True values were obtained via Monte Carlo with \(10^6\) independent participants to approximate the empirical transition-specific survival functions under each arm and then obtain the true value for RMT-IF estimands. Details are provided in the Web Appendix C.1. We compared five estimators: four variants of the proposed doubly robust estimator (depending on different model specifications) and the nonparametric estimator in Mao.\cite{mao2023restricted} For brevity, we use ``o'' for the outcome regression \(P\{T_i^{q,(a)} \ge t \mid Z_{i1},Z_{i2}\}\) and ``c'' for the censoring survival \(K_c^{(a)}(t \mid Z_{i1},Z_{i2})\), where ``1'' denotes correct specification and ``0'' denotes misspecification. Misspecification is induced by omitting \(Z_{i1}Z_{i2}\) interaction term. Specifically:
\begin{enumerate}[i.]
\item (o1c1): Both outcome and censoring models are correctly specified.
\item (o1c0): Outcome model is correctly specified; censoring model is mis-specified (drops \(Z_{i1}Z_{i2}\)).
\item (o0c1): Outcome model is mis-specified (drops \(Z_{i1}Z_{i2}\)); censoring model is correctly specified.
\item (o0c0): Both outcome and censoring models are mis-specified (\(Z_{i1}Z_{i2}\)).
\item (RMT-IF): The nonparametric RMT-IF estimator proposed by Mao \cite{mao2023restricted}.
\end{enumerate}
Variance estimates were computed using the group jackknife described in Section \ref{jk_var_iid}, with \(K=100\) groups for all the above five models. Across 1{,}000 Monte Carlo replications, performance was evaluated using the following metrics: percentage absolute relative bias \(\mathrm{PBias}=\left|\mathbb{E}[\widehat\Delta^{\text{rmt-if}}(t)]-\Delta^{\text{rmt-if}}(t)\right|/\left|\Delta^{\text{rmt-if}}(t)\right|\); average empirical standard error (AESE), defined as the mean of the jackknife standard errors over replications; Monte Carlo standard deviation (MCSD), given by the empirical standard deviation of the point estimates; and empirical coverage probability (CP) of nominal 95\% confidence intervals.

Table \ref{sim_win_iid} summarizes the results for the estimands \(\xi^{(1)}(t)\), \(\xi^{(0)}(t)\), and their contrast \(\Delta^{\text{rmt-if}}(t)\) at times \(t=1,1.5,2\). Across all model specifications (o1c1, o1c0, o0c1, o0c0), the proposed estimators display very small percentage bias, generally below 1\% even at the latest time point. The AESE closely tracks the MCSD, and the empirical coverage probabilities remain close to the nominal 95\% level. This indicates that the proposed estimators generally perform well, even when only one of the outcome or censoring models is correctly specified.  When both models are misspecified (o0c0), there is a mild increase in percentage bias, especially at later time points (up to about 0.8\% for \(\Delta^{\text{rmt-if}}(t)\) at \(t=2\)), and the coverage probabilities show a slight decline. Nonetheless, the performance remains satisfactory, with coverage still around 95\%.  In contrast, the nonparametric RMT-IF estimator exhibits substantially larger bias, which grows as time increases: from about 4\% for \(\Delta^{\text{rmt-if}}(t)\) at \(t=1\) to nearly 9.5\% at \(t=2\). Moreover, its coverage probabilities deteriorate severely with time, dropping from 92\% at \(t=1\) to below 70\% at \(t=2\). This is because the nonparametric RMT-IF estimator is not designed to address covariate-dependent censoring.

\begin{table}[!ht]
\caption{Simulation results for the estimands \(\xi^{(1)}(t)\) and \(\xi^{(0)}(t)\), and their contrast \(\Delta^{\text{rmt-if}}(t)\), estimated under different working models for the outcome and censoring processes, together with the competing RMT-IF estimator under randomized controlled trials (RCTs) with \(N=1000\) individuals. Results are shown at evaluation times \(t=\{1,1.5,2\}\). Reported metrics include percent bias (PBias, in \(\%\)), AESE (average empirical standard error), MCSD (Monte Carlo standard deviation of the point estimates), and CP (empirical coverage probability of the nominal 95\% confidence interval).}
\label{sim_win_iid}
\centering
\resizebox{\textwidth}{!}{%
\begin{tabular}{llrrrrrrrrrrrr}
\toprule
\multicolumn{2}{c}{} &
\multicolumn{4}{c}{$\xi^{(1)}(t)$} &
\multicolumn{4}{c}{$\xi^{(0)}(t)$} &
\multicolumn{4}{c}{$\Delta^{\text{rmt-if}}$} \\
\cmidrule(lr){3-6}\cmidrule(lr){7-10}\cmidrule(lr){11-14}
Method & $t$ &
PBias (\%) & AESE & MCSD & CP &
PBias (\%) & AESE & MCSD & CP &
PBias (\%) & AESE & MCSD & CP \\
\midrule
o1c1 & 1   & 0.437 & 0.025 & 0.025 & 0.938 & 0.596 & 0.028 & 0.029 & 0.918 & 1.100 & 0.012 & 0.012 & 0.943 \\
     & 1.5 & 0.566 & 0.039 & 0.040 & 0.927 & 0.737 & 0.044 & 0.046 & 0.911 & 0.700 & 0.019 & 0.020 & 0.947 \\
     & 2.0 & 0.672 & 0.054 & 0.055 & 0.926 & 0.854 & 0.059 & 0.062 & 0.905 & 0.435 & 0.027 & 0.028 & 0.948 \\
o1c0 & 1   & 0.431 & 0.025 & 0.025 & 0.939 & 0.593 & 0.028 & 0.029 & 0.920 & 1.131 & 0.012 & 0.012 & 0.943 \\
     & 1.5 & 0.561 & 0.039 & 0.040 & 0.926 & 0.735 & 0.044 & 0.046 & 0.921 & 0.725 & 0.019 & 0.020 & 0.947 \\
     & 2.0 & 0.668 & 0.054 & 0.055 & 0.923 & 0.852 & 0.059 & 0.062 & 0.927 & 0.454 & 0.027 & 0.027 & 0.949 \\
o0c1 & 1   & 0.416 & 0.025 & 0.025 & 0.939 & 0.576 & 0.028 & 0.029 & 0.924 & 1.127 & 0.012 & 0.013 & 0.944 \\
     & 1.5 & 0.499 & 0.039 & 0.040 & 0.936 & 0.675 & 0.044 & 0.046 & 0.916 & 0.795 & 0.020 & 0.020 & 0.947 \\
     & 2.0 & 0.547 & 0.053 & 0.055 & 0.921 & 0.736 & 0.059 & 0.062 & 0.920 & 0.605 & 0.027 & 0.027 & 0.943 \\
o0c0 & 1   & 0.350 & 0.025 & 0.025 & 0.945 & 0.509 & 0.028 & 0.029 & 0.925 & 1.184 & 0.012 & 0.013 & 0.942 \\
     & 1.5 & 0.400 & 0.039 & 0.040 & 0.940 & 0.579 & 0.044 & 0.046 & 0.920 & 0.921 & 0.020 & 0.020 & 0.946 \\
     & 2.0 & 0.416 & 0.054 & 0.055 & 0.934 & 0.614 & 0.059 & 0.062 & 0.913 & 0.790 & 0.027 & 0.027 & 0.945 \\
RMT-IF & 1   & 1.677 & 0.026 & 0.026 & 0.659 & 1.435 & 0.031 & 0.032 & 0.829 & 4.006 & 0.016 & 0.016 & 0.921 \\
       & 1.5 & 2.657 & 0.041 & 0.042 & 0.437 & 2.074 & 0.048 & 0.049 & 0.753 & 6.950 & 0.027 & 0.027 & 0.832 \\
       & 2.0 & 3.592 & 0.057 & 0.059 & 0.279 & 2.624 & 0.064 & 0.066 & 0.685 & 9.490 & 0.038 & 0.037 & 0.696 \\
\bottomrule
\end{tabular}%
}
\end{table}

%\section{Empirical validation for CRT} \label{simu:crt}

\subsection{Simulations under cluster randomization}
We next consider a two-arm CRT, where the binary treatment indicator \(A_i\) for cluster \(i\) is \(A_i = 1\) if the cluster is randomized to the experimental intervention and \(A_i = 0\) otherwise, with \(P(A_i = 1) = P(A_i = 0) = 0.5\). The study included \(M = 60\) clusters, where the cluster size \(N_i\) followed a discrete uniform distribution from \(\{10,,\dots, 90\}\). Each cluster was simulated with two cluster-level covariates and two individual-level covariates. Specifically, the covariate vector \(\bm{V}_{ij}\) for individual \(j\) in cluster \(i\) includes \(W_{i1}\) and \(W_{i2}\) for the first and second cluster-level covariates, and \(Z_{ij1}\) and \(Z_{ij2}\) for the first and second individual-level covariates. Among these, \(W_{i1} \sim \text{Bernoulli}(p = 0.5)\), \(W_{i2} \sim \text{Normal}(\mu = N_i / 50, \text{sd} = 1.5)\), \(Z_{ij1} \sim \text{Normal}(\mu = \log(N_i) / 5, \text{sd} = 1)\), and \(Z_{ij2} \sim \text{Bernoulli}(p = 0.5)\). Both \(W_{i2}\) and \(Z_{ij1}\) depend on the cluster size \(N_i\).

We consider three states for event times, where the third state is an absorbing state. The event times in all states are generated using a frailty Cox model. The frailty \(B_i\) follows a \(\text{Gamma}(2, 2)\) distribution with Kendall’s \(\tau = 0.2\) for \(A_i = 1\) and a \(\text{Gamma}(4.5, 4.5)\) distribution with Kendall’s \(\tau = 0.1\) for \(A_i = 0\). The potential event time \(\widetilde{T}_{ij}^{1,(a)}\) in state 1 for \(A_i = a_i\) is generated from the hazard function:
\[
\lambda_1^{(a)} (t|\bm{V}_{ij}) = \frac{N_i}{100} \left\{0.01 - 0.005(1-a_i)\right\} B_i \exp\left\{-a_i + W_{i1} + 2W_{i2} + Z_{ij1} - 0.6Z_{ij2} + Z_{ij1}\frac{N_i}{\mathbb E[N_i]} + \frac{N_i}{\mathbb{E}[N_i]} \right\}.
\]
The event time in state 2 is defined as \(\widetilde{T}_{ij}^{2,(a)} = \widetilde{T}_{ij}^{1,(a)} + u_{ij}\), where the gap time \(u_{ij}\) is generated from the hazard function:
\[
\lambda_u^{(a)} (u|\bm{V}_{ij}) = \frac{N_i}{100} \left\{2 - (1-a_i)\right\} B_i \exp\left\{-0.5a_i - W_{i1} + W_{i2} + 2Z_{ij1} - Z_{ij2} - Z_{ij1}\frac{N_i}{\mathbb E[N_i]} +   \frac{N_i}{\mathbb E[N_i]} \right\}.
\]
The event time in the third (absorbing) state, \(T_{ij}^{3,(a)}\), is generated from the hazard function:
\[
\lambda_3^{(a)} (t|\bm{V}_{ij}) = \frac{N_i}{100} \left\{0.08 - 0.04(1-a_i)\right\} B_i \exp\left\{-2a_i -W_{i1} + W_{i2} + 2Z_{ij1} - Z_{ij2} -Z_{ij1}\frac{N_i}{\mathbb E[N_i]} + \frac{N_i}{\mathbb E[N_i]}  \right\}.
\]
Censoring times \(C_{ij}\) were generated independently from a frailty model, with frailty \(R_i \sim \text{Gamma}(9.5, 9.5)\) and Kendall’s \(\tau = 0.05\). The censoring hazard function is \(
h^{(a)}(t|\bm{V}_{ij}) = 0.13 R_i \exp\left\{a+ 0.5W_{i1} -0.5W_{i2} -0.8Z_{ij1} +0.5Z_{ij2}\right\}\).
The event times are defined as \(T_{ij}^{1,(a)} = \min\{\widetilde{T}_{ij}^{1,(a)}, \widetilde{T}_{ij}^{2,(a)}, T_{ij}^{3,(a)}\}\) and \(T_{ij}^{2,(a)} = \min\{\widetilde{T}_{ij}^{2,(a)}, T_{ij}^{3,(a)}\}\). The observed time is \(U_{ij}^{q,(a)} = T_{ij}^{q,(a)} \wedge C_{ij}\) for \(q \in \{1, 2, 3\}\), with status indicator \(\delta_{ij} = 1\) if \(T_{ij}^{1,(a)} \leq C_{ij}\), \(\delta_{ij} = 2\) if \(T_{ij}^{2,(a)} \leq C_{ij}\), and \(\delta_{ij} = 3\) if \(T_{ij}^{3,(a)} \leq C_{ij}\). The overall censoring rate was controlled by adjusting \(h_0\) and parameters \(\alpha_1, \ldots, \alpha_4\) to achieve approximately 50\% censoring.

Our target estimands include \(\xi_{C}^{(a)}(t)\), \(\xi_{I}^{(a)}(t)\), and their restricted mean time in favor times \(\Delta_{C}^{\text{rmt-if}}(t)\) and \(\Delta_{I}^{\text{rmt-if}}(t)\) on the difference scale. The true values of these estimands were determined through Monte Carlo simulations, where a large sample of \(10^6\) clusters was generated to compute the empirical survival functions at both the cluster and individual levels (Web Appendix C.2). We evaluated nine estimation strategies in total: eight based on our proposed doubly robust approach (depending on model specifications) and a nonparametric estimator adapted from Mao \cite{mao2023restricted} to address clustering. The specifications are as follows:
\begin{itemize}
\item (marginal-o1c1): proposed estimator with correctly specified censoring model \(K_c^{(a)}(t\mid \bm{V}_{ij})\) and outcome model \(P(T_{ij}^{(a)} \geq t \mid \bm{V}_{ij})\) using marginal Cox models.
\item (marginal-o1c0): proposed estimator with mis-specified censoring model, where \(Z_{ij1}\) is omitted from \(K_c^{(a)}(t\mid \bm{V}_i)\), and correctly specified outcome model using marginal Cox models.
\item (marginal-o0c1): proposed estimator with correctly specified censoring model but mis-specified outcome model, where both \(N_i\) and the interaction term \(Z_{ij1}N_i\) are omitted from \(P(T_{ij}^{(a)} \geq t \mid \bm{V}_{ij})\), using marginal Cox models.
\item (marginal-o0c0): proposed estimator with both censoring and outcome models mis-specified as defined above, using marginal Cox models.
\item (frailty-o1c1): proposed estimator with correctly specified censoring and outcome models using frailty Cox models with Gamma frailty consistent with the data-generating process.
\item (frailty-o1c0): proposed estimator with mis-specified censoring model (omit \(Z_{ij1}\)) and correctly specified outcome model using frailty Cox models with Gamma frailty.
\item (frailty-o0c1): proposed estimator with correctly specified censoring model and mis-specified outcome model (omit \(N_i\) and \(Z_{ij1}N_i\)) using frailty Cox models with Gamma frailty.
\item (frailty-o0c0): proposed estimator with both censoring and outcome models mis-specified as defined above, using frailty Cox models with Gamma frailty.
\item (rmt-if): the rmt-if estimator of Mao\cite{mao2023restricted}, modified for clustered data in CRTs. Specifically, survival probabilities are estimated nonparametrically and participants are weighted inversely by cluster size (\(1/N_i\)) for cluster-level estimators and equally by 1 for individual-level estimators.
\end{itemize}

We conducted 1,000 Monte Carlo iterations with iteration-indexed random seeds for reproducibility. Results are summarized in terms of percentage absolute bias (PBias), empirical standard error from the cluster jackknife (AESE), Monte Carlo standard deviation (MCSD), and empirical coverage probability (CP) of nominal 95\% confidence intervals.

%\subsection{Study results}

Table \ref{sim_win_crt_cluster} summarizes the simulation results for the cluster-level estimands \(\xi_C^{(1)}(t)\), \(\xi_C^{(0)}(t)\), and their contrast \(\Delta_C^{\text{rmt-if}}(t)\) at evaluation times \(t=\{1,1.5,2\}\). When both the censoring model \(K_c^{(a)}(t\mid \bm V_{ij})\) and the outcome model \(P(T_{ij}^{(a)} \geq t \mid \bm V_{ij})\) are correctly specified, the proposed estimators under both marginal and frailty Cox models are approximately unbiased, with percent bias well below 1\% for the component estimands and less than 1.5\% for the contrast. In these settings, the empirical coverage probabilities are close to the nominal 95\% level, confirming the adequate performance of the cluster jackknife variance estimator. When only one of the two models is correctly specified, performance remains similar to the correctly specified case. In particular, under marginal-o1c0 and marginal-o0c1, bias remains small, AESE continues to track MCSD, and coverage remains near nominal. Frailty-based estimators show the same pattern, with slightly larger percent bias but comparable efficiency. In contrast, when both working models are mis-specified (marginal-o0c0 and frailty-o0c0), the doubly robust estimators show slightly higher percent bias, especially for the contrast \(\Delta_C^{\text{rmt-if}}(t)\) at earlier time points, and both AESE and MCSD increase compared to the correctly specified settings. Nevertheless, coverage probabilities remain around 95\%. The adapted nonparametric RMT-IF estimator shows a different pattern. It suffers from substantial bias that increases with \(t\) for both the component estimands and the contrast, with PBias for \(\Delta_C^{\text{rmt-if}}(t)\) exceeding 17\% at \(t=1\). Its AESE and MCSD are also larger, and coverage for the component estimands is far below nominal (42\%-67\%), although coverage for the contrast happens to be closer to 95\%. Results for the corresponding individual-level estimands \(\xi_I^{(a)}(t)\) and \(\Delta_I^{\text{rmt-if}}(t)\) are reported in Web Appendix D, with similar trends in bias, efficiency, and coverage.

\begin{table}[!ht]
\caption{Simulation results for the cluster-level estimands \(\xi_C^{(1)}(t)\) and \(\xi_C^{(0)}(t)\), and their contrast \(\Delta_C^{\text{rmt-if}}(t)\), estimated under different working models for the outcome and censoring processes, together with the competing RMT-IF estimator under cluster randomized trial (CRT) with \(M=50\) individuals. Results are shown at evaluation times \(t=\{1,1.5,2\}\). Reported metrics include percent bias (PBias, in \(\%\)), AESE (average empirical standard error), MCSD (Monte Carlo standard deviation of the point estimates), and CP (empirical coverage probability of the nominal 95\% confidence interval).}
\label{sim_win_crt_cluster}
\centering
\resizebox{\textwidth}{!}{%
\begin{tabular}{llrrrrrrrrrrrr}
\toprule
\multicolumn{2}{c}{} &
\multicolumn{4}{c}{$\xi_C^{(1)}(t)$} &
\multicolumn{4}{c}{$\xi_C^{(0)}(t)$} &
\multicolumn{4}{c}{$\Delta_C^{\text{rmt-if}}$} \\
\cmidrule(lr){3-6}\cmidrule(lr){7-10}\cmidrule(lr){11-14}
Method & $t$ &
PBias (\%) & AESE & MCSD & CP &
PBias (\%) & AESE & MCSD & CP &
PBias (\%) & AESE & MCSD & CP \\
\midrule
\multicolumn{14}{c}{\textbf{Doubly robust estimator (Marginal Cox)}} \\
\addlinespace[2pt]
marginal-o1c1 & 1   & 0.400 & 0.120 & 0.122 & 0.942 & 0.465 & 0.127 & 0.129 & 0.943 & 1.359 & 0.025 & 0.023 & 0.952 \\
              & 1.5 & 0.413 & 0.185 & 0.188 & 0.942 & 0.441 & 0.197 & 0.200 & 0.944 & 0.191 & 0.039 & 0.037 & 0.954 \\
              & 2   & 0.432 & 0.249 & 0.253 & 0.944 & 0.411 & 0.266 & 0.271 & 0.945 & 0.815 & 0.055 & 0.052 & 0.955 \\
marginal-o1c0 & 1   & 0.381 & 0.120 & 0.122 & 0.942 & 0.451 & 0.127 & 0.129 & 0.942 & 1.517 & 0.024 & 0.023 & 0.955 \\
              & 1.5 & 0.388 & 0.185 & 0.188 & 0.942 & 0.428 & 0.197 & 0.200 & 0.944 & 0.470 & 0.039 & 0.037 & 0.953 \\
              & 2   & 0.394 & 0.250 & 0.254 & 0.944 & 0.396 & 0.267 & 0.271 & 0.946 & 0.351 & 0.054 & 0.052 & 0.956 \\
marginal-o0c1 & 1   & 0.383 & 0.121 & 0.123 & 0.935 & 0.479 & 0.129 & 0.131 & 0.936 & 2.223 & 0.044 & 0.042 & 0.953 \\
              & 1.5 & 0.417 & 0.187 & 0.190 & 0.940 & 0.503 & 0.200 & 0.203 & 0.937 & 1.403 & 0.068 & 0.064 & 0.956 \\
              & 2   & 0.456 & 0.252 & 0.256 & 0.942 & 0.515 & 0.270 & 0.275 & 0.936 & 0.603 & 0.093 & 0.088 & 0.955 \\
marginal-o0c0 & 1   & 0.281 & 0.122 & 0.123 & 0.936 & 0.405 & 0.129 & 0.131 & 0.937 & 3.080 & 0.044 & 0.041 & 0.952 \\
              & 1.5 & 0.264 & 0.187 & 0.190 & 0.940 & 0.393 & 0.200 & 0.203 & 0.936 & 2.500 & 0.068 & 0.064 & 0.957 \\
              & 2   & 0.247 & 0.253 & 0.257 & 0.940 & 0.370 & 0.271 & 0.276 & 0.939 & 1.965 & 0.092 & 0.088 & 0.955 \\
\addlinespace[3pt]
\multicolumn{14}{c}{\textbf{Doubly robust estimator (Frailty Cox)}} \\
\addlinespace[2pt]
frailty-o1c1  & 1   & 0.769 & 0.120 & 0.119 & 0.955 & 0.865 & 0.127 & 0.128 & 0.942 & 1.815 & 0.025 & 0.023 & 0.954 \\
              & 1.5 & 0.798 & 0.184 & 0.184 & 0.954 & 0.856 & 0.197 & 0.199 & 0.943 & 0.440 & 0.039 & 0.037 & 0.950 \\
              & 2   & 0.820 & 0.249 & 0.248 & 0.952 & 0.837 & 0.267 & 0.270 & 0.940 & 0.527 & 0.055 & 0.052 & 0.950 \\
frailty-o1c0  & 1   & 0.844 & 0.120 & 0.119 & 0.955 & 0.912 & 0.128 & 0.128 & 0.942 & 1.007 & 0.025 & 0.023 & 0.953 \\
              & 1.5 & 0.918 & 0.185 & 0.184 & 0.954 & 0.943 & 0.197 & 0.199 & 0.943 & 0.381 & 0.039 & 0.037 & 0.944 \\
              & 2   & 0.987 & 0.249 & 0.248 & 0.950 & 0.967 & 0.267 & 0.270 & 0.943 & 1.357 & 0.055 & 0.052 & 0.948 \\
frailty-o0c1  & 1   & 0.797 & 0.121 & 0.121 & 0.944 & 0.919 & 0.129 & 0.130 & 0.946 & 2.506 & 0.044 & 0.041 & 0.962 \\
              & 1.5 & 0.842 & 0.187 & 0.188 & 0.945 & 0.961 & 0.200 & 0.202 & 0.939 & 1.690 & 0.068 & 0.063 & 0.967 \\
              & 2   & 0.881 & 0.252 & 0.253 & 0.946 & 0.987 & 0.270 & 0.274 & 0.938 & 1.041 & 0.093 & 0.086 & 0.964 \\
frailty-o0c0  & 1   & 0.911 & 0.121 & 0.121 & 0.945 & 0.994 & 0.129 & 0.130 & 0.947 & 1.314 & 0.044 & 0.041 & 0.964 \\
              & 1.5 & 1.032 & 0.187 & 0.186 & 0.949 & 1.097 & 0.200 & 0.201 & 0.943 & 0.346 & 0.069 & 0.064 & 0.966 \\
              & 2   & 1.148 & 0.252 & 0.252 & 0.947 & 1.190 & 0.270 & 0.273 & 0.942 & 0.386 & 0.094 & 0.087 & 0.966 \\
\addlinespace[3pt]
\multicolumn{14}{c}{\textbf{Non-parametric RMT-IF}} \\
\addlinespace[2pt]
RMT-IF        & 1   & 10.998 & 0.106 & 0.102 & 0.421 & 12.052 & 0.123 & 0.118 & 0.496 & 17.548 & 0.076 & 0.072 & 0.948 \\
              & 1.5 & 11.982 & 0.170 & 0.164 & 0.442 & 13.463 & 0.199 & 0.190 & 0.513 & 19.627 & 0.121 & 0.115 & 0.945 \\
              & 2   & 12.474 & 0.236 & 0.228 & 0.488 & 14.331 & 0.276 & 0.264 & 0.543 & 20.988 & 0.167 & 0.160 & 0.946 \\
\bottomrule
\end{tabular}%
}
\end{table}

\section{Two Illustrative Data Applications}
\label{data_ex}

\subsection{Systolic Blood Pressure Intervention Trial (SPRINT)}

The Systolic Blood Pressure Intervention Trial (SPRINT) was a large, multicenter, individually randomized trial that compared two systolic blood-pressure targets, an intensive target of $<\!120$\,mmHg versus a standard target of $<\!140$\,mmHg in adults aged $\ge 50$ years at elevated cardiovascular risk but without diabetes or prior stroke.\cite{sprint2015randomized,ambrosius2014design} SPRINT enrolled and randomized $N=9{,}361$ participants across 102 U.S. clinics. The primary composite outcome was the first occurrence of nonfatal myocardial infarction (MI), non-MI acute coronary syndrome (ACS), nonfatal stroke, nonfatal heart failure (HF), or cardiovascular (CVD) death. The original analyses using Cox proportional hazards models for time to first occurrence of the composite endpoint and for all-cause mortality found that intensive SBP targeting significantly reduced both outcomes (primary composite with HR$=0.75$, 95\% CI $0.64$-$0.89$; all-cause mortality with HR$=0.73$, 95\% CI $0.60$-$0.90$). However, each occurrence of these nonfatal events was prospectively recorded, and participants could experience multiple events over follow-up, giving an event sequence rather than a single event. A time-to-first analysis thus has notable limitations; for example, it discards information on recurrences and their timing. Motivated by these considerations, we incorporate the occurrence of all nonfatal events (HF, MI, non--MI ACS, stroke) by viewing the disease course as a progressive recurrent process: any first nonfatal event moves a participant from the initial state to a higher-morbidity state, each additional new nonfatal event constitutes a transition to the next, more severe state, and CVD death is modeled as an absorbing state. After removing the missing observations with missing values, we included $9{,}244$ participants in our analysis. 
%, with $4{,}616$ participants in the standard-treatment strategy and \(4{,}628\) participants in the intensive-treatment strategy. 
The total number of non-fatal events occurred in the standard-treatment strategy is 457 (3.13 event rate per 100 person year), along with 72 fatal events, compared with the 343 (2.34 event rate per 100 person year) non-fatal events with 45 fatal events in the intensive-treatment strategy.

We consider the event history as a 6-state process in which participants may experience up to 5 distinct non-fatal events, with death treated as an absorbing state (state 6). We winsorize the non-fatal event count at 5; that is, participants with more than five non-fatal events prior to death are recoded as having five events, because such high-recurrence were rare (see Web Appendix F).
 The causal estimand of interest is the restricted mean time in favor, \(\Delta^{\text{rmt-if}}(t)\), defined as the between-arm difference in restricted mean time in favor up to time \(t\) for the two systolic blood-pressure targets. We analyzed SPRINT using our proposed approach with state-specific Cox models for the outcome \(P\{T_i^{q,(a)} > t \mid \bm Z_i\}\), and the censoring \(K_c^{(a)}(t \mid \bm Z_i)\), alongside the nonparametric estimator of Mao \cite{mao2023restricted}. Covariate adjustment included participation in the SPRINT-MIND cognitive substudy, participation in the SPRINT-MIND MRI substudy, participation in the health-related quality-of-life substudy, chronic kidney disease at baseline, age, an indicator for age \(\ge 75\) years, and baseline cardiovascular disease history. The same covariates were used in both the state-specific outcome models and the censoring model. %To assess covariate-dependent censoring, we 
Interestingly, when we fit a Cox model for the censoring hazard including these covariates, all were statistically significant at the 0.05 level except baseline cardiovascular disease history, suggesting some evidence against completely independent censoring. 
%, supporting covariate adjustment in \(K_c^{(a)}(t \mid \bm Z_i)\) for our proposed method. 
The 95\% of confidence interval was calculated using the group jackknife method described in Section \ref{jk_var_iid} with \(K=100\). We chose \(K=100\) for consistency with the simulation study, and to balance computational burden with numerical stability of the group jackknife. Repeating the analysis with \(K \in \{50, 200\}\) gave similar results.

\begin{figure}
    \centering
    \includegraphics[width=0.6\linewidth]{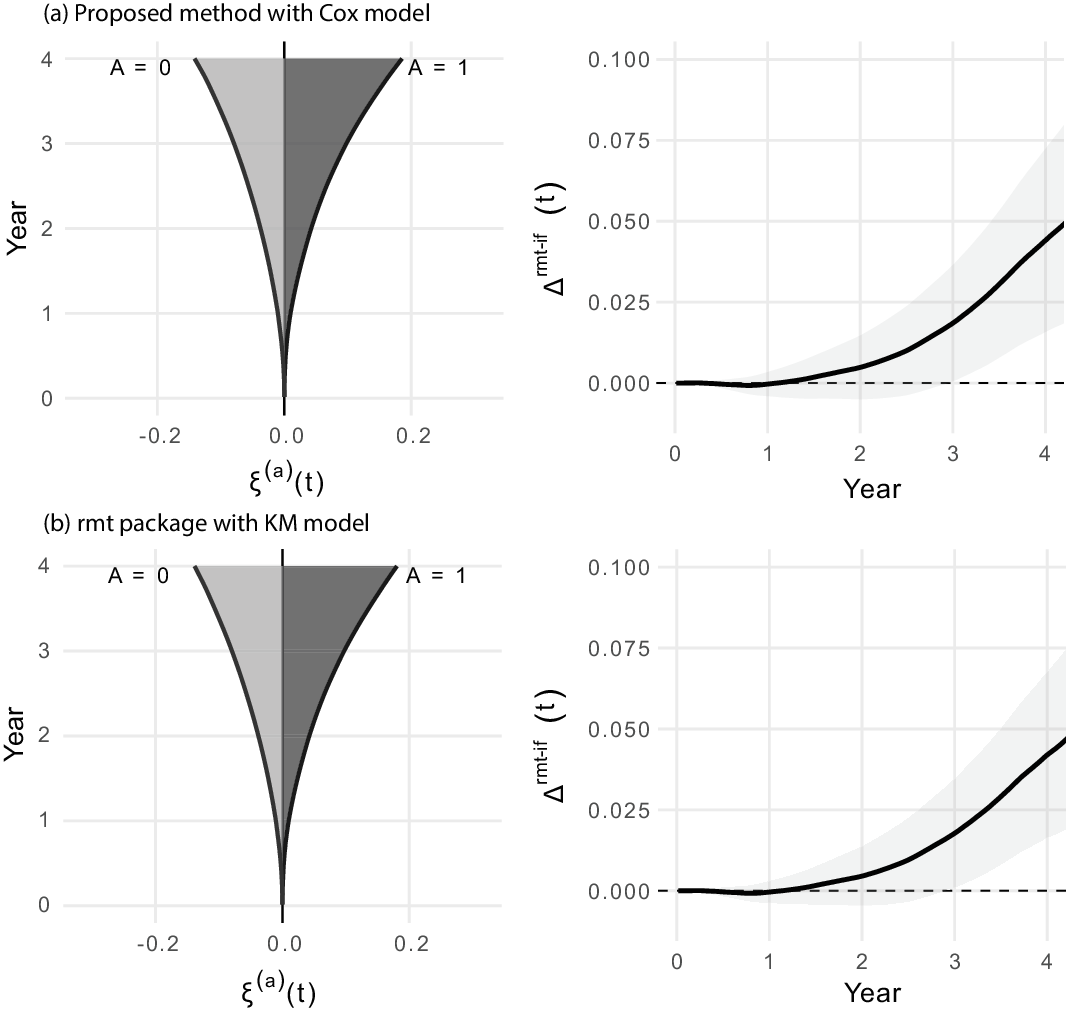}
    \caption{Estimated RMT-IF, $\xi^{(a)}(t)$ (left panels), and the causal effect, $\Delta^{\text{rmt-if}}(t)$ (right panels), from the SPRINT data. For each method, the left panel displays the bouquet plot of the arm-specific RMT-IF, and the right panel shows the corresponding estimated causal effect. Panel (a) reports results from our proposed doubly robust method. Panel (b) reports the nonparametric estimator using the package from Mao \cite{mao2023restricted}.}
\label{fig:sprint}
\end{figure}

Figure \ref{fig:sprint} presents the estimated causal effect $\Delta^{\text{rmt-if}}(t)$ up to 4 years for the two systolic blood-pressure targets. Panel (a) shows our proposed doubly robust estimator, and Panel (b) shows the nonparametric estimator of Mao \cite{mao2023restricted}, each with pointwise confidence bands based on a $t$ distribution ($\text{df}=99$). The left panels present bouquet plots of the arm-specific RMT-IF estimates \(\xi^{(a)}(t)\), adapted from the visualization tool of Mao \cite{mao2023restricted}, to illustrate differences between the treatment and control arms as follow-up progresses. At a given time \(t\), the width of the bouquet summarizes the average time over \([0,t]\) during which two patients drawn from the two arms would be in discordant outcome states, providing an intuitive indication of the information available for between-arm comparison. The direction and extent of skewness reflect the direction and magnitude of the overall treatment effect. The right panels present the estimated causal effect \(\Delta^{\text{rmt-if}}(t)\) over time. Across follow-up, the two approaches give similar trajectories with confidence bands of comparable width. This similarity is largely owing to the modest covariate effects in both the outcome and censoring processes, where the estimated log-hazard ratios range approximately from $-0.17$ to $0.45$. However, a key advantage of our proposed method is its doubly robust property, which provides protection against bias arising from misspecification of either the outcome or censoring model. Accordingly, our interpretation focuses on the doubly robust estimator, for which the estimated treatment effect $\Delta^{\text{rmt-if}}(t)$ remains near zero through about 2.5 years, shows modest departures from zero at intermediate times, and becomes statistically different from zero beginning around year 3; thereafter, the effect persists through year 4 with a gradually increasing magnitude. %Both estimators concur on the onset of significance and the overall time profile of the effect.

\subsection{Strategies to Reduce Injuries and Develop Confidence in Elders (STRIDE) Trial}
The Strategies to Reduce Injuries and Develop Confidence in Elders (STRIDE) is a pragmatic, cluster-randomized trial designed to evaluate the effectiveness of a multifactorial intervention aimed at preventing fall injuries among older adults at risk of fall.\cite{Bhasin2018,Bhasin2020}  A total of 5451 community-dwelling adults aged 70 years or older were enrolled from 86 primary care across 10 health care systems, with practices randomized equally into either the intervention group (43 practices) or the control group (43 practices). The trial's primary outcome was time to first serious fall-related injury with maximum follow-up time 44 months. Of the total participants, 2,802 were assigned to the intervention group and 2,649 to the control group. During the study, 3,002 participants reported no fall injuries, while 1,372 experienced one fall injury, 594 experienced two, 258 experienced three, and 225 experienced four or more fall injuries. Additionally, 232 participants died during the follow-up period. Cluster sizes varied widely across practices, ranging from 10 to 199 participants, with a mean cluster size of 63 and a coefficient of variation of 0.517; the non-trivial cluster size variation could have suggested potentially informative cluster size and hence motivates us to consider our methods.

We consider the recurrent serious fall injury as a progressive multi-state outcome. In STRIDE, the maximum number of reported fall injuries is 29. But because only a few single individuals reported number of recurrences beyond 9 (see Web Appendix Section F), we winsorize the data up to 9 recurrences to ensure the feasibility of estimating state-specific survival functions. Thus, in this illustration, we consider 10 states, with death as an absorbing state (state 10). Each reported fall injury represented a distinct state, and transitions occurred to subsequent states with additional reported fall injuries until death. The causal estimands of interest includes cluster-level and individual-level restricted mean time in favor,  \(\Delta_C^{\text{rmt-if}}(t)\) and \(\Delta_I^{\text{rmt-if}}(t)\), which represent the differences in restricted mean time in favor between the intervention and control groups at the cluster and individual levels, respectively. The data were analyzed using 1) our proposed method under marginal Cox working models and 2) the RMT-IF method proposed by Mao \cite{mao2023restricted} as a possible comparison. In the RMT-IF method, we modify the non-parametric estimation of the survival function by applying a weight of \(1/N_i\) for cluster-level estimators and an equal weight of 1 for individual-level estimators.  In the proposed method, we adjust the covariates, including race, gender, and number of chronic conditions in both censoring model and outcome model. In addition, the cluster size is included as a covariate in outcome model. Standard errors for the estimators and 95\% confidence intervals are calculated using the cluster jackknife method described in Section \ref{jk_var_crt}.

\begin{figure}
    \centering
    \includegraphics[width=1\linewidth]{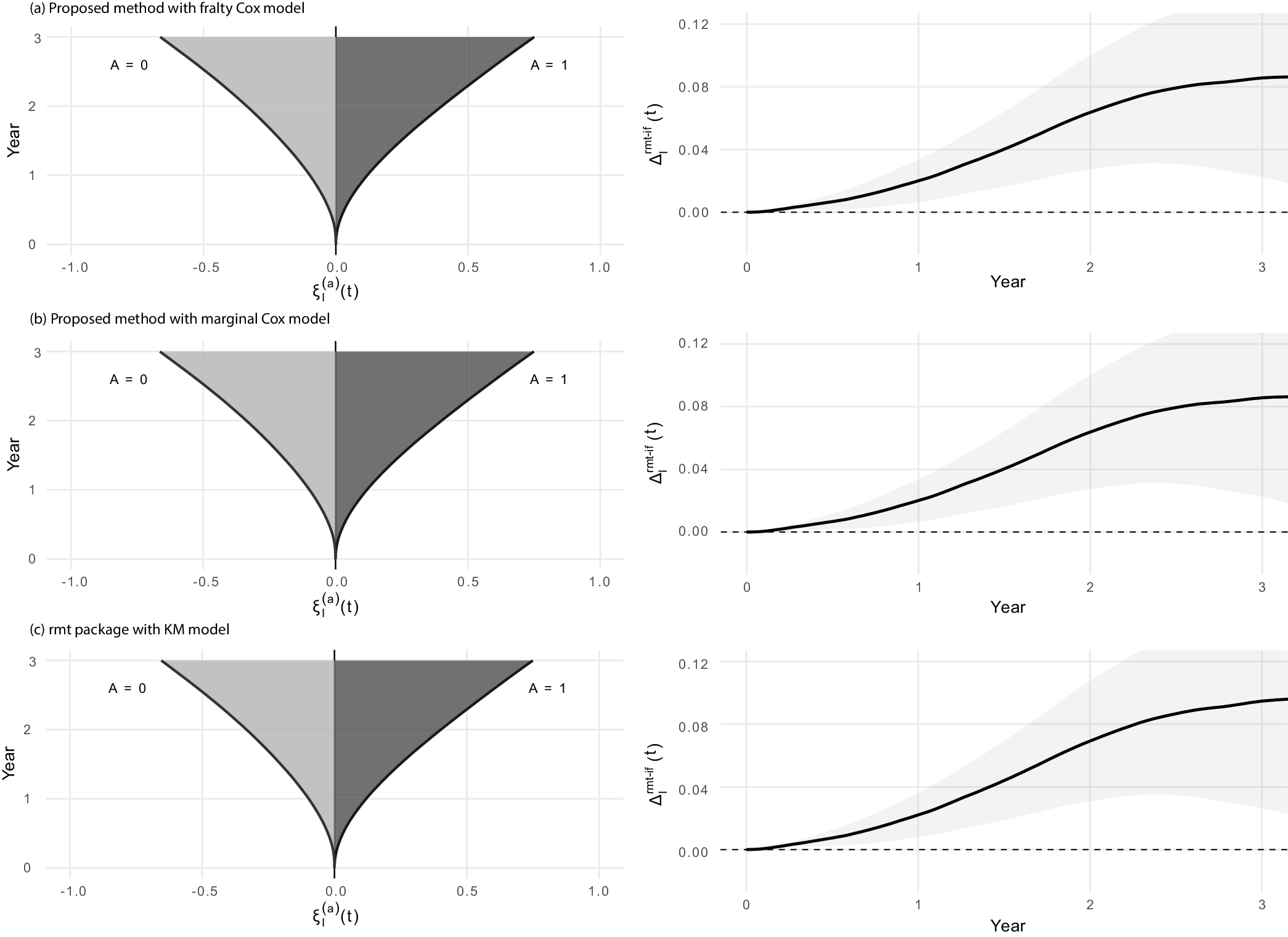}
    \caption{Estimated individual-level RMT-IF, $\xi_I^{(a)}(t)$ (left panels), and the causal effect, $\Delta_I^{\text{rmt-if}}(t)$ (right panels), from the STRIDE data. For each method, the left panel displays the bouquet plot of the arm-specific individual-level RMT-IF, and the right panel shows the corresponding estimated causal effect. Panel (a) reports results from the proposed doubly robust method using stage-specific frailty Cox models for $P\{T_{ij}^{q,(a)} > t \mid \bm Z_i\}$ and for the censoring distribution $K_c^{(a)}(t \mid \bm V_{ij})$. Panel (b) reports results from the proposed doubly robust method using stage-specific marginal Cox models for $P\{T_{ij}^{q,(a)} > t \mid \bm V_i\}$ and for the censoring distribution. Panel (c) reports the nonparametric estimator of Mao \cite{mao2023restricted} using Kaplan--Meier estimators for $P\{T_{ij}^{q,(a)} > t \mid \bm V_i\}$ and $K_c^{(a)}(t \mid \bm V_i)$ with equal weight for each individual.}
\label{fig:stride_i}
\end{figure}

Figure \ref{fig:stride_i} presents the STRIDE analysis of the individual-level arm-specific RMT-IF estimands and their causal effect using three estimation methods with different models for the censoring and outcome processes. The left panel shows the bouquet plot of the treatment-specific trajectories for $\xi_I^{(a)}(t)$, and the right panel shows the corresponding contrast $\Delta_I^{\text{rmt-if}}(t)$, with pointwise 95\% confidence bands based on a $t$ distribution with $\text{df}=85$. At the individual level, all three methods produce highly similar trajectories for both $\xi_I^{(a)}(t)$ and $\Delta_I^{\text{rmt-if}}(t)$. The estimated treatment effect departs from the null and remains above zero over mid-to-late follow-up, with the confidence bands excluding zero, indicating a statistically significant, time-accumulating treatment benefit. Notably, the RMT-IF estimator proposed by Mao \cite{mao2023restricted} is closely aligned with the proposed methods at the individual level, despite some loss of efficiency due to the lack of covariate adjustment. For example, the proposed method with marginal Cox models for both the outcome and censoring processes achieves approximately 7.5\%, 10.7\%, and 10.9\% efficiency gains relative to the nonparametric RMT-IF estimator at 1, 1.5, and 2 years, respectively. This is consistent with our simulation results, where covariate adjustment for the outcome and censoring processes led to efficiency gains for the doubly robust method compared with non-adjusted nonparametric estimators.

The corresponding cluster-level results are provided in Web Appendix E. In contrast to the individual-level analysis, the proposed doubly robust estimators produce consistent trajectories for $\xi_C^{(a)}(t)$ and $\Delta_C^{\text{rmt-if}}(t)$ across all combinations of outcome and censoring models. However, the nonparametric estimator of Mao \cite{mao2023restricted}, even after reweighting each individual by $1/N_i$ to target cluster-level state-specific survival functions, produces RMT-IF trajectories that differ substantially from the doubly robust estimators. This pattern is consistent with our simulation study, where the cluster-level RMT-IF estimator exhibited substantial bias in the presence of informative cluster size. Thus, our primary interpretation of the cluster-level treatment effect is based on the doubly robust estimator, which indicates effects that are smaller in magnitude than the corresponding individual-level effects and generally not statistically significant over most of the follow-up. The discrepancies between the cluster-level and individual-level trajectories suggest the presence of informative cluster size and are consistent with our previous findings for the time-to-first injury outcome, where cluster-level and individual-level estimators also lead to different treatment effects in both magnitude and significance.\cite{fang2025estimands} These results highlight the importance of carefully choosing between cluster-level and individual-level estimands in CRT with multistate survival outcomes. In STRIDE, the active intervention is a nurse-delivered, individualized fall-prevention program implemented directly at the patient level, which makes the individual-level treatment effect a particularly natural and policy-relevant target and explains more pronounced and statistically significant effects observed at the individual level.

 \section{Discussion}
\label{discussion}

In this work, we develop an estimand-aligned framework for studying the treatment effects with progressive multi-state survival outcomes, focusing on the restricted mean time in favor (RMT-IF) as a marginal estimand. The RMT-IF generalizes the restricted mean survival time (RMST) to accommodate prioritized, time-dependent outcomes, providing a patient-centered and interpretable summary of the net survival time gained under treatment versus control. \cite{mao2023restricted,mao2024dissecting} Unlike conventional hazard ratios, which are non-collapsible and may lack causal interpretations even under randomization, \cite{aalen2015does} the RMT-IF expresses treatment benefit on an absolute time scale and remains valid when the proportional hazards assumption fails. This estimand also aligns with the recent emphasis on estimand clarity and pre-specification articulated in the ICH E9(R1) addendum.\cite{lewis1999statistical} Our main contribution is to provide a doubly robust estimator for the RMT-IF that remains consistent if either the censoring or the outcome regression model is correctly specified. The estimator combines augmented inverse-probability weighting with stage-specific outcome modeling, providing valid inference without relying on parametric multi-state intensity assumptions. Compared to the existing nonparametric RMT-IF estimator,\cite{mao2023restricted} our estimator incorporates baseline covariates (via survival outcome modeling) for improved efficiency and mitigates potential bias (via inverse probability of censoring weighting) from covariate-dependent censoring. %By expressing the estimand as a functional of marginal survival distributions rather than transition-specific hazards, we preserve interpretability while achieving robustness, offering a practical and theoretically sound framework for multistate survival analyses. 

We further extend the RMT-IF estimand and its doubly robust estimation to CRTs, where the choice of estimand critically depends on the unit of inference.\cite{kahan2023estimands} This work builds on a growing literature on estimands in CRTs. Li et al.\cite{li2025model} introduced cluster-level and individual-level treatment effect estimands for non-survival outcomes, including continuous, binary, and count data, and proposed model-robust standardization estimators that remain valid under broad outcome model misspecification. Fang et al.\cite{fang2025estimands} extended this framework to right-censored survival outcomes with single-state time-to-event endpoints and developed doubly robust estimators for two natural survival-based targets defined through the survival function and the RMST. In comparison, our focus is on progressive multistate survival processes in CRTs, where outcomes evolve through ordered states over time. We define RMT-IF estimands at the cluster level through pairwise comparison of treatment and control regimes along the multistate trajectory, which captures the cumulative net benefit of treatment across clinically ordered states. Within this framework, we formalized two natural targets, the cluster-average treatment effect (c-ATE) and the individual-average treatment effect (i-ATE), but both are now defined on the pairwise comparison regimes with natural differences how each cluster-pair or individual-pair is weighted. We develop doubly robust estimators for both c-ATE and i-ATE that accommodate covariate-dependent censoring and intracluster correlation by combining stage-specific outcome regression models with censoring models. These estimators remain consistent if either the outcome regression model or the censoring model is correctly specified. In the special case of completely independent censoring, they reduce to model-robust standardization estimators that retain consistency even under outcome model misspecification, therefore recovering and extending the robustness properties established in Li et al.\cite{li2025model}. 

Several directions for further methodological development emerge from this work. The doubly robust approach can naturally incorporate flexible machine-learning estimators for nuisance functions, including random survival forests, \cite{ishwaran2008random} gradient boosting, \cite{chen2013gradient,morris2020survboost} and neural network-based survival models,\cite{hao2021deep} which may potentially improve estimation efficiency while capturing nonlinear and high-dimensional relationships. Extending the current estimator through cross-fitting and targeted learning frameworks \cite{westling2024inference} would further ensure asymptotic validity even when using complex learners, but is not a trivial task and requires separate developments. Another second potential direction is time-global inference. Constructing uniform confidence bands and integrated tests for $\Delta^{\text{rmt-if}}(t)$ would allow rigorous evaluation of evolving treatment effects over time. %Expanding the framework to stepped-wedge and other longitudinal CRTs, or to settings that permit limited interference between clusters, would also increase its practical relevance. As pragmatic trials increasingly collect high-frequency and high-dimensional longitudinal data, integrating the RMT-IF framework with dynamic treatment-regime estimation may enable individualized evaluation of sustained interventions. 
%Finally, developing sample-size and power calculation methods for the RMT-IF in CRTs, building on recent planning tools for win-based and RMST-type endpoints,\cite{mao2023study} will facilitate its implementation in trial design and improve alignment between design, estimand, and analysis.
Finally, although sample size methods for RMT-IF have been recently discussed in individually randomized trials by Mao,\cite{mao2023study} corresponding study planning tools for cluster-randomized designs remain largely undeveloped. Recent work by Fang et al.\cite{fang2025sample} has established new sample size procedures for win-based statistics in CRTs, highlighting the critical role of the rank intracluster correlation coefficient as a central quantity. Expanding their study design approach to accommodate RMT-IF with progressive multi-state outcomes would further support design of CRTs with multistate endpoints.

\section*{Acknowledgement}
Research in this article was supported by the National Heart, Lung, and Blood Institute under the Award Number 1R01HL178513-01. The statements presented in this article are solely the responsibility of the authors and do not necessarily represent the official views of the National Institutes of Health. The STRIDE study was funded primarily by the Patient Centered Outcomes Research Institute (PCORI\textsuperscript{\textregistered}), with additional support from the National Institute on Aging (NIA) at NIH. Funding is provided and the award managed through a cooperative agreement (5U01AG048270) between the NIA and the Brigham and Women’s Hospital. %The authors thank Professor Peter Peduzzi for help in accessing the STRIDE data. %The authors also thank the Associate Editor and two anonymous reviewers for providing constructive comments that have improved the quality of this article.

\section*{Supplementary Material}

The supplementary material includes technical derivations and web tables referenced in the article.

\section*{Data Availability Statement}

An R package implementing our method is available at CRAN \url{https://cran.csail.mit.edu/web/packages/DRsurvCRT/}, and a tutorial for using this package is provided in Web Appendix F. The STRIDE data can be obtained via \url{https://agingresearchbiobank.nia.nih.gov/studies/}, which is publicly available.

\clearpage
\newpage
\printbibliography
%%% and comment out the ``thebibliography'' section.

\newpage
\appendix

\section{Proof of Proposition 2} \label{supp:proof_dr}

\begin{proof}
We establish double robustness for the stage- and arm-specific survival estimator $\widehat S^{q,(a)}(t)$ in (2). The result for $\widehat\Delta^{\mathrm{rmt\text{-}if}}(t)$ then follows by the plug-in representations. Let $q\in\{1,\dots,Q+1\}$ and $t\ge 0$, and consider $a=1$ (the case $a=0$ is analogous). Define $\widetilde N_i(u)=\mathbb{I}(C_i\le u)$ and the corresponding censoring martingale increment
\(d\widetilde M_i(u\mid \bm Z_i)= d\widetilde N_i(u)-\lambda_c^{(1)}(u\mid \bm Z_i)\,\mathbb{I}(C_i\ge u)\,du\), where $\lambda_c^{(1)}(u\mid \bm Z_i)=-\{d\log K_c^{(1)}(u\mid \bm Z_i)\}/du$ and $K_c^{(1)}(u\mid \bm Z_i)=\mathbb{P}(C_i\ge u\mid A_i=1,\bm Z_i)$. By Bai et al.\ \cite{bai2013doubly}, for all $t$,
\begin{align}
\int_{0}^{t}\frac{d\widetilde N_i(u)-\lambda_c^{(1)}(u\mid \bm Z_i)\mathbb{I}(C_i\ge u)\,du}{K_c^{(1)}(u\mid \bm Z_i)}=
\int_{0}^{t}\frac{d\widetilde M_i(u\mid \bm Z_i)}{K_c^{(1)}(u\mid \bm Z_i)}=1-\frac{\mathbb{I}(C_i\ge t)}{K_c^{(1)}(t\mid \bm Z_i)}.
\label{eq:dr_pre_irt}
\end{align}

Assume $\widehat P(T_i^{q,(1)}\ge t\mid \bm Z_i)\stackrel{p}{\to}P^{*}(T_i^{q,(1)}\ge t\mid \bm Z_i)$ and $\widehat K_c^{(1)}(t\mid \bm Z_i)\stackrel{p}{\to}K_c^{(1),*}(t\mid \bm Z_i)$, and recall $\pi^{(1)}=\mathbb{P}(A_i=1)$ is known by design. Using (2) and $\mathbb{I}(U_i^q\ge t)=\mathbb{I}(T_i^q\ge t)\mathbb{I}(C_i\ge t)$, we can write
\begin{align}
\widehat S^{q,(1)}(t)
=&\ \frac{1}{N}\sum_{i=1}^N\Bigg[
\frac{A_i\,\mathbb{I}(T_i^q\ge t)\mathbb{I}(C_i\ge t)}{\pi^{(1)}\,\widehat K_c^{(1)}(t\mid \bm Z_i)}
-\Bigg\{\frac{A_i-\pi^{(1)}}{\pi^{(1)}}\Bigg\}\widehat P(T_i^{q,(1)}\ge t\mid \bm Z_i) \nonumber\\
&\hspace{2.1cm}
+\frac{A_i}{\pi^{(1)}}\int_{0}^{t}
\frac{d\widehat M_{c,i}^{q,(1)}(u\mid \bm Z_i)}{\widehat K_c^{(1)}(u\mid \bm Z_i)}
\frac{\widehat P(T_i^{q,(1)}\ge t\mid \bm Z_i)}{\widehat P(T_i^{q,(1)}\ge u\mid \bm Z_i)}
\Bigg]  \nonumber\\
=&\ \frac{1}{N}\sum_{i=1}^N\Bigg[
\frac{A_i\,\mathbb{I}(T_i^{q,(1)}\ge t)\mathbb{I}(C_i^{(1)}\ge t)}{\pi^{(1)}\,K_c^{(1),*}(t\mid \bm Z_i)}
-\Bigg\{\frac{A_i-\pi^{(1)}}{\pi^{(1)}}\Bigg\}P^{*}(T_i^{q,(1)}\ge t\mid \bm Z_i) \nonumber\\
&\hspace{2.1cm}
+\frac{A_i}{\pi^{(1)}}\int_{0}^{t}
\frac{d M_{c,i}^{q,(1)}(u\mid \bm Z_i)}{K_c^{(1),*}(u\mid \bm Z_i)}
\frac{P^{*}(T_i^{q,(1)}\ge t\mid \bm Z_i)}{P^{*}(T_i^{q,(1)}\ge u\mid \bm Z_i)}
\Bigg] + o_p(1),
\label{eq:S_start_irt}
\end{align}
where the second equality uses SUTVA and the replacement of nuisance estimators by their probability limits. By \eqref{eq:dr_pre_irt} with $K_c^{(1),*}$ in place of $K_c^{(1)}$,
\[
\frac{\mathbb{I}(C_i^{(1)}\ge t)}{K_c^{(1),*}(t\mid \bm Z_i)}=1-\int_0^t \frac{d\widetilde M_i(u\mid \bm Z_i)}{K_c^{(1),*}(u\mid \bm Z_i)}.
\]
Substituting this identity into \eqref{eq:S_start_irt} and rearranging yields the decomposition
\begin{align}
\widehat S^{q,(1)}(t)
=&\ \frac{1}{N}\sum_{i=1}^N\Bigg[
\mathbb{I}(T_i^{q,(1)}\ge t)
+\underbrace{\Bigg\{\frac{A_i-\pi^{(1)}}{\pi^{(1)}}\Bigg\}
\Bigg\{\mathbb{I}(T_i^{q,(1)}\ge t)-P^{*}(T_i^{q,(1)}\ge t\mid \bm Z_i)\Bigg\}}_{(8)} \nonumber\\
&\hspace{1.25cm}
+\underbrace{\frac{A_i}{\pi^{(1)}}\int_{0}^{t}
\frac{d\widetilde M_i(u\mid \bm Z_i)}{K_c^{(1),*}(u\mid \bm Z_i)}
\Bigg\{\frac{\mathbb{I}(T_i^{q,(1)}\ge u)\,P^{*}(T_i^{q,(1)}\ge t\mid \bm Z_i)}
{P^{*}(T_i^{q,(1)}\ge u\mid \bm Z_i)}
-\mathbb{I}(T_i^{q,(1)}\ge t)\Bigg\}}_{(9)}
\Bigg] + o_p(1).
\label{eq:S_decomp_irt}
\end{align}

If the outcome regression is correctly specified, i.e., $P^{*}(T_i^{q,(1)}\ge t\mid \bm Z_i)=P(T_i^{q,(1)}\ge t\mid \bm Z_i)$, then $\mathbb{E}\{(8)\}=0$ because $\mathbb{E}(A_i-\pi^{(1)}\mid \bm Z_i)=0$ by design and $\mathbb{E}\{\mathbb{I}(T_i^{q,(1)}\ge t)\mid \bm Z_i\}=P(T_i^{q,(1)}\ge t\mid \bm Z_i)$. If the censoring model is correctly specified, i.e., $K_c^{(1),*}(u\mid \bm Z_i)=K_c^{(1)}(u\mid \bm Z_i)$, then by Assumption (3) the process $\widetilde M_i(u\mid \bm Z_i)$ is a martingale increment and the integrand in (9) is predictable and integrable, implying $\mathbb{E}\{(9)\}=0$. In either case, taking expectations in \eqref{eq:S_decomp_irt} gives $\mathbb{E}\{\widehat S^{q,(1)}(t)\}=S^{q,(1)}(t)+o(1)$. Since $\{O_i\}_{i=1}^N$ are i.i.d.\ and the summands in \eqref{eq:S_decomp_irt} have finite variance under standard positivity conditions, a law of large numbers argument yields $\widehat S^{q,(1)}(t)\stackrel{p}{\to}S^{q,(1)}(t)$ if either the censoring model $K_c^{(1)}(\cdot\mid\bm Z)$ or the outcome regression $P(T^{q,(1)}\ge \cdot\mid \bm Z)$ is correctly specified. The same argument applies for $a=0$ and for each $q$. Finally, by Proposition (1), each $\xi^{q,(a)}(t)$ and $\Delta^{q,\mathrm{rmt\text{-}if}}(t)$ is a smooth functional of $\{S^{q,(a)}(\cdot),S^{q+1,(a)}(\cdot)\}$. Therefore, the corresponding plug-in estimators obtained by replacing $S$ with $\widehat S$ are consistent under the same union model, and summing over $q$ yields consistency of $\widehat\Delta^{\mathrm{rmt\text{-}if}}(t)=\sum_{q=1}^{Q+1}\widehat\Delta^{q,\mathrm{rmt\text{-}if}}(t)$.
\end{proof}

The proof for the CRT estimators is analogous, where clusters are independent units and replacing the individual summand in \eqref{eq:S_decomp_irt} by the either cluster-level or individual level summand, the same mean-zero arguments for terms (8) and (9) under Assumption (4) establish double robustness of $\widehat S_C^{q,(a)}(t)$ (and hence $\widehat S_I^{q,(a)}(t)$), and the result for $\widehat\Delta_C^{\mathrm{rmt\text{-}if}}(t)$ and $\widehat\Delta_I^{\mathrm{rmt\text{-}if}}(t)$ follows by plug-in.

\section{Proof of Proposition 3}
\begin{proof}
Recall that the state-wise CRT estimands and their arm-specific components are 
\[\Delta_{C}^{q,\mathrm{rmt\text{-}if}}(t) = \xi_{C}^{q,(1)}(t) -\xi_{C}^{q,(0)}(t),  \qquad  \xi_{C}^{q,(a)}(t)   = \mathbb{E} \left[ \frac{1}{N_i N_k}\sum_{j=1}^{N_i}\sum_{l=1}^{N_k}  \mathcal{W}^q\{Y^{(a)}_{ij},Y^{(1-a)}_{kl}\}(t)\right],
\]
and the cluster-level stage-specific survival functions \(S_{C}^{q,(a)}(t)=\mathbb{E}\!\left\{\frac{1}{N_i}\sum_{j=1}^{N_i} \mathbb{I}\big(T_{ij}^{q,(a)}\ge t\big)\right\}\). By the definition of the state-specific win time,
\[\mathcal{W}^q\{Y^{(a)}_{ij},Y^{(1-a)}_{kl}\}(t)=\int_0^t \mathbb{I}\{Y^{(a)}_{ij}(u)<q,\;Y^{(1-a)}_{kl}(u)=q\}\,du,\]
and linearity of the integral and expectation, we have
\begin{align} \label{eq_clus_rmt}
\xi_{C}^{q,(a)}(t)&=\mathbb{E} \left[ \frac{1}{N_i N_k}\sum_{j=1}^{N_i}\sum_{l=1}^{N_k} \mathcal{W}^q\{Y^{(a)}_{ij},Y^{(1-a)}_{kl}\}(t)\right] \nonumber \\
&=\mathbb{E} \left[ \frac{1}{N_i N_k}\sum_{j=1}^{N_i}\sum_{l=1}^{N_k} \int_0^t \mathbb{I}\{Y^{(a)}_{ij}(u)<q,\;Y^{(1-a)}_{kl}(u)=q\}\,du \right] \nonumber \\
&=\int_0^t \mathbb{E} \left[ \frac{1}{N_i N_k}\sum_{j=1}^{N_i}\sum_{l=1}^{N_k} \mathbb{I}\{Y^{(a)}_{ij}(u)<q,\;Y^{(1-a)}_{kl}(u)=q\}\right]\,du.    
\end{align}
By the progressive multistate structure,
\[
\{Y^{(a)}_{ij}(u)<q\} \iff \{T_{ij}^{q,(a)}>u\}, \qquad \{Y^{(1-a)}_{kl}(u)=q\} \iff \{T_{kl}^{q,(1-a)}\le u < T_{kl}^{q+1,(1-a)}\}.
\]
Hence
\[\mathbb{I}\{Y^{(a)}_{ij}(u)<q,\;Y^{(1-a)}_{kl}(u)=q\}=\mathbb{I}\big(T_{ij}^{q,(a)}>u\big)\,\Big\{\mathbb{I}\big(T_{kl}^{q+1,(1-a)}>u\big) -\mathbb{I}\big(T_{kl}^{q,(1-a)}>u\big)\Big\}.
\]
Substituting into \eqref{eq_clus_rmt}, and by cluster randomization, we have 
\begin{align*}
\xi_{C}^{q,(a)}(t) &= \int_0^t  \mathbb{E} \left[ \frac{1}{N_i N_k}\sum_{j=1}^{N_i}\sum_{l=1}^{N_k}  \mathbb{I}\big(T_{ij}^{q,(a)}>u\big)\, \Big\{\mathbb{I}\big(T_{kl}^{q+1,(1-a)}>u\big)-\mathbb{I}\big(T_{kl}^{q,(1-a)}>u\big)\Big\}
\right]\,du \\
& = \int_0^t\mathbb{E} \left[ \frac{1}{N_i N_k}\sum_{j=1}^{N_i}\sum_{l=1}^{N_k} \mathbb{I}\big(T_{ij}^{q,(a)}>u\big)\,\Big\{\mathbb{I}\big(T_{kl}^{q+1,(1-a)}>u\big) -\mathbb{I}\big(T_{kl}^{q,(1-a)}>u\big)\Big\}\right]\\
&= \int_0^t \mathbb{E}\left[\frac{1}{N_i}\sum_{j=1}^{N_i}\mathbb{I}\big(T_{ij}^{q,(a)}>u\big)\right]\,\mathbb{E}\left[\frac{1}{N_k}\sum_{l=1}^{N_k} \Big\{\mathbb{I}\big(T_{kl}^{q+1,(1-a)}>u\big) -\mathbb{I}\big(T_{kl}^{q,(1-a)}>u\big)\Big\}\right].\\
& =\int_0^t 
S_{C}^{q,(a)}(u)\,
\Big\{S_{C}^{q+1,(1-a)}(u) - S_{C}^{q,(1-a)}(u)\Big\}\,du.
\end{align*}

For the individual-level estimand, recall that
\[\Delta_{I}^{q,\mathrm{rmt\text{-}if}}(t) = \xi_{I}^{q,(1)}(t) -\xi_{I}^{q,(0)}(t),\quad \xi_{I}^{q,(a)}(t)  =\frac{\mathbb{E} \left[ \sum_{j=1}^{N_i}\sum_{l=1}^{N_k} \mathcal{W}^q\{Y^{(a)}_{ij},Y^{(1-a)}_{kl}\}(t)\right]}{\mathbb{E}(N_i N_k)},\]
with \(S_{I}^{q,(a)}(t)=\frac{\mathbb{E}\!\left\{\sum_{j=1}^{N_i}\mathbb{I}\big(T_{ij}^{q,(a)}\ge t\big)\right\}}{\mathbb{E}(N_i)}\). Through the definition above and cluster randomization, we have
\begin{align*}
    \xi_{I}^{q,(a)}(t)&=\frac{\mathbb{E} \left[ \sum_{j=1}^{N_i}\sum_{l=1}^{N_k} \mathcal{W}^q\{Y^{(a)}_{ij},Y^{(1-a)}_{kl}\}(t)\right]}{\mathbb{E}(N_i N_k)}\\
&=\frac{\mathbb{E} \left[ \sum_{j=1}^{N_i}\sum_{l=1}^{N_k} \int_0^t \mathbb{I}\{Y^{(a)}_{ij}(u)<q,\;Y^{(1-a)}_{kl}(u)=q\}\,du \right]}{\mathbb{E}(N_i N_k)}\\
&=\int_0^t \frac{\mathbb{E} \left[ \sum_{j=1}^{N_i}\sum_{l=1}^{N_k} \mathbb{I}\{Y^{(a)}_{ij}(u)<q,\;Y^{(1-a)}_{kl}(u)=q\}\right]}{\mathbb{E}(N_i N_k)}\,du \\
&=\int_0^t \frac{\mathbb{E} \left[ \sum_{j=1}^{N_i}\sum_{l=1}^{N_k} \mathbb{I}\big(T_{ij}^{q,(a)}>u\big)\,\Big\{\mathbb{I}\big(T_{kl}^{q+1,(1-a)}>u\big) -\mathbb{I}\big(T_{kl}^{q,(1-a)}>u\big)\Big\}\right]}{\mathbb{E}(N_i N_k)}\,du\\
& = \int_0^t \frac{\mathbb{E}\!\left[\sum_{j=1}^{N_i}\mathbb{I}\big(T_{ij}^{q,(a)}>u\big)\right]\,\mathbb{E}\!\left[\sum_{l=1}^{N_k}\Big\{\mathbb{I}\big(T_{kl}^{q+1,(1-a)}>u\big)-\mathbb{I}\big(T_{kl}^{q,(1-a)}>u\big)\Big\}\right]}{\mathbb{E}(N_i N_k)}\,du\\
&=\int_0^t \left[\frac{\mathbb{E}\!\left\{\sum_{j=1}^{N_i}\mathbb{I}\big(T_{ij}^{q,(a)}>u\big)\right\}}{\mathbb{E}(N_i)}\right]\left(\frac{\mathbb{E}\!\left[\sum_{l=1}^{N_k}\Big\{\mathbb{I}\big(T_{kl}^{q+1,(1-a)}>u\big) -\mathbb{I}\big(T_{kl}^{q,(1-a)}>u\big)\Big\}\right]}{\mathbb{E}(N_k)}\right)\,du\\
& = \int_0^t 
S_{I}^{q,(a)}(u)\,
\Big\{S_{I}^{q+1,(1-a)}(u) - S_{I}^{q,(1-a)}(u)\Big\}\,du.
\end{align*}

\end{proof}

\section{True values of RMT-IF in the simulation studies under IRT and CRT}

\subsection{True estimand values under IRT}

Under the simulation design in Section 4.1, we consider a three-state progressive process with states \(q=1,2,3\), where state 3 is terminal. For state 1, conditional on \((A_i=a, Z_{i1},Z_{i2})\), we have \(\widetilde T_{i}^{1,(a)} \sim \text{Exp}\big(\mu^{(a)}_1\big)\) with rate \(\mu^{(a)}_{i1} = \lambda^{(a)}_{i1}(t \mid Z_{i1},Z_{i2})\). For state 2, the gap time \(u_i \sim \text{Exp}\big(\mu^{(a)}_{i2}\big)\) with \(\mu^{(a)}_{i2}= \lambda^{(a)}_u(u \mid Z_{i1},Z_{i2})\) and for state 3, \(T_{i}^{3,(a)} \sim \text{Exp}\big(\mu^{(a)}_{i3}\big)\) with rate \(\mu_{i3}^{(a)} = \lambda^{(a)}_3(t \mid Z_{i1},Z_{i2})\), where the transition time \(\widetilde T_{i}^{2,(a)} = \widetilde T_{i}^{1,(a)} + u_i\). The stage-specific potential transition times are then
\[
T_{i}^{1,(a)} = \min\{\widetilde T_{i}^{1,(a)},\,\widetilde T_{i}^{2,(a)},\,T_{i}^{3,(a)}\}
= \min\{\widetilde T_{i}^{1,(a)},\,T_{i}^{3,(a)}\},\qquad T_{i}^{2,(a)} = \min\{\widetilde T_{i}^{2,(a)},\,T_{i}^{3,(a)}\},\qquad T_{i}^{3,(a)} = T_{i}^{3,(a)},
\]
so that \(T_{i}^{1,(a)} \le T_{i}^{2,(a)} \le T_{i}^{3,(a)}\). Under the above exponential distribution, direct calculations give
\begin{align*}
S^{3,(a)}(t \mid \mu_{i3}^{(a)})&= \Pr\{T_{i}^{3,(a)} > t \mid A_i=a, Z_{i1},Z_{i2}\}= \exp\{-\mu^{(a)}_{i3} t\},\\
S^{1,(a)}(t \mid Z_{i1},Z_{i2}) &= \Pr\{\widetilde T_{i}^{1,(a)} > t,\, T_{i}^{3,(a)} > t \mid A_i=a, Z_{i1},Z_{i2}\} = \exp\{-(\mu^{(a)}_{i1} + \mu^{(a)}_{i3}) t\},\\
S^{2,(a)}(t \mid Z_{i1},Z_{i2}) &= \Pr\{\min(\widetilde T_{i}^{1,(a)}+u_i,\,T_{i}^{3,(a)}) > t \mid A_i=a, Z_{i1},Z_{i2}\} \\
&= \frac{\mu^{(a)}_{i2}}{\mu^{(a)}_{i2} - \mu^{(a)}_{i1}}\exp\{-(\mu^{(a)}_{i1}+\mu^{(a)}_{i3})t\}- \frac{\mu^{(a)}_{i1}}{\mu^{(a)}_{i2} - \mu^{(a)}_{i1}}\exp\{-(\mu^{(a)}_{i2}+\mu^{(a)}_{i3})t\},
\end{align*}
The marginal stage-specific survival functions entering the RMT-IF estimand are
\[
S^{q,(a)}(t) = \Pr\{T_{i}^{q,(a)} > t\} = \mathbb{E}_{Z_{i1},Z_{i2}}\big\{ S^{q,(a)}(t \mid Z_{i1},Z_{i2}) \big\},\qquad q=1,2,3,\; a\in\{0,1\},
\]
where the expectation is taken over the baseline covariate distribution. These integrals are evaluated numerically by Monte Carlo, where we generate a large sample with \(N= 10^{6}\) from their data-generating distribution, compute \(S^{q,(a)}(t \mid Z_{i1},Z_{i2})\) for each draw and a fine grid of time points, and take average to obtain \(S^{q,(a)}(t)\).  Given the marginal survival functions \(\{S_{q,(a)}(t): q=1,2,3;\, a=0,1\}\), the stage-specific win function for arm \(a\) at severity level \(q\) is
\[
\xi^{q,(a)}(t) = \mathbb{E}\big[ \mathcal{W}_q\{Y^{(a)}_i,Y^{(1-a)}_j\}(t) \big] = \int_0^t S^{q,(a)}(u)\,\big\{ S^{q+1,(1-a)}(u) - S^{q,(1-a)}(u) \big\}\,du,
\]
which are computed by numerically integrating the above expressions over the fine time grid used via the trapezoidal rule.

\subsection{True estimand values under CRT}

Under the CRT simulation design in Section 4.2, we again consider the three-state progressive process with states \(q=1,2,3\), where state 3 is terminal, but now the stage-specific potential transition times \(T_{ij}^{q,(a)}\) are generated from the multistate frailty Cox model for each arm \(a\in\{0,1\}\), cluster \(i\), and individual \(j\). In this setting, the marginal stage-specific survival functions entering the cluster-level and individual-level RMT-IF estimands are defined as
\[S_{C}^{q,(a)}(t) = \mathbb{E}\!\left[\,\frac{1}{N_i}\sum_{j=1}^{N_i} \mathbb{I}\{T_{ij}^{q,(a)} > t\}\right], \qquad q=1,2,3,\; a\in\{0,1\},
\]
\[
S_{I}^{q,(a)}(t)  = \frac{\mathbb{E}\!\left[\sum_{j=1}^{N_i} \mathbb{I}\{T_{ij}^{q,(a)} > t\}\right]}{\mathbb{E}(N_i)},
\qquad q=1,2,3,\; a\in\{0,1\},
\]
where \(\mathbb{E}(\cdot)\) is taken with respect to the joint distribution of the baseline covariates, cluster sizes, and frailties under the CRT data-generating mechanism. Because the multistate frailty Cox model does not give closed-form expressions for these survival functions, we evaluate them numerically by Monte Carlo through generate a large super-population of \(M = 10^{6}\) clusters from the CRT data-generating distribution, simulate \(\{T_{ij}^{q,(a)}\}\) for all stages and arms on a fine grid of time points, and approximate
\[
 S_{ C}^{q,(a)}(t)
= \frac{1}{M}\sum_{i=1}^{M}\left\{\frac{1}{N_i}\sum_{j=1}^{N_i} S_{C}^{q,(a)}(t\mid \bm{V}_{ij},N_{i},B_i)\right\},
\qquad
 S_{I}^{q,(a)}(t)
= \frac{\sum_{i=1}^{M}\sum_{j=1}^{N_i} S_{C}^{q,(a)}(t\mid \bm{V}_{ij},N_{i},B_i)}{\sum_{i=1}^{M} N_i}.
\]
Given the marginal stage-specific survival functions \(\{S_{\mathrm C}^{q,(a)}(t), S_{\mathrm I}^{q,(a)}(t): q=1,2,3;\, a=0,1\}\), the stage-specific cluster-level and individual-level win functions for arm \(a\) are
\begin{align*}
    \xi_{C}^{q,(a)}(t) & = \int_0^t S_{\mathrm C}^{q,(a)}(u)\,\big\{ S_{C}^{q+1,(1-a)}(u) - S_{C}^{q,(1-a)}(u) \big\}\,du \\
   \xi_{I}^{q,(a)}(t) & =  \int_0^t S_{\mathrm I}^{q,(a)}(u)\,\big\{ S_{I}^{q+1,(1-a)}(u) - S_{I}^{q,(1-a)}(u) \big\}\,du,
\end{align*}
which are computed by numerically integrating the above expressions over the fine time grid via the trapezoidal rule.

\section{Additional simulation results for individual RMT-IR}\label{additional_results}

Table~\ref{sim_win_crt_individual} summarizes the finite-sample performance of the proposed doubly robust RMT-IF estimators and the non-parametric RMT-IF estimator for \(\xi_I^{(1)}(t)\), \(\xi_I^{(0)}(t)\), and \(\Delta_I^{\text{rmt-if}}(t)\) under the CRT design with \(M=50\), showing uniformly small percent bias, close agreement between AESE and MCSD, and empirical coverage probabilities close to the nominal 95\% level for all methods and time points, with the doubly robust procedures additionally achieving smaller standard deviation for the treatment contrast \(\Delta_I^{\text{rmt-if}}(t)\) compared with the non-parametric estimator.

\begin{table}[!ht]
\caption{Simulation results for the individual-level estimands \(\xi_I^{(1)}(t)\) and \(\xi_I^{(0)}(t)\), and their contrast \(\Delta_I^{\text{rmt-if}}(t)\), estimated under different working models for the outcome and censoring processes, together with the competing RMT-IF estimator under cluster randomized trial (CRT) with \(M=50\) individuals. Results are shown at evaluation times \(t=\{1,1.5,2\}\). Reported metrics include percent bias (PBias, in \(\%\)), AESE (average empirical standard error), MCSD (Monte Carlo standard deviation of the point estimates), and CP (empirical coverage probability of the nominal 95\% confidence interval).}
\label{sim_win_crt_individual}
\centering
\resizebox{\textwidth}{!}{%
\begin{tabular}{llrrrrrrrrrrrr}
\toprule
\multicolumn{2}{c}{} &
\multicolumn{4}{c}{$\xi_I^{(1)}(t)$} &
\multicolumn{4}{c}{$\xi_I^{(0)}(t)$} &
\multicolumn{4}{c}{$\Delta^{\text{rmt-if}}$} \\
\cmidrule(lr){3-6}\cmidrule(lr){7-10}\cmidrule(lr){11-14}
Method & $t$ &
PBias (\%) & AESE & MCSD & CP &
PBias (\%) & AESE & MCSD & CP &
PBias (\%) & AESE & MCSD & CP \\
\midrule
\multicolumn{14}{c}{\textbf{Doubly robust estimator (Marginal Cox)}} \\
\addlinespace[2pt]
marginal-o1c1 & 1   & 0.410 & 0.133 & 0.135 & 0.935 & 0.546 & 0.143 & 0.144 & 0.937 & 2.204 & 0.031 & 0.030 & 0.949 \\
              & 1.5 & 0.442 & 0.200 & 0.202 & 0.936 & 0.556 & 0.214 & 0.217 & 0.936 & 1.276 & 0.049 & 0.046 & 0.954 \\
              & 2   & 0.476 & 0.264 & 0.267 & 0.938 & 0.553 & 0.284 & 0.286 & 0.938 & 0.483 & 0.067 & 0.063 & 0.956 \\
marginal-o1c0 & 1   & 0.435 & 0.134 & 0.135 & 0.934 & 0.566 & 0.143 & 0.144 & 0.939 & 2.090 & 0.031 & 0.029 & 0.948 \\
              & 1.5 & 0.488 & 0.200 & 0.202 & 0.934 & 0.599 & 0.214 & 0.216 & 0.936 & 1.179 & 0.048 & 0.046 & 0.953 \\
              & 2   & 0.540 & 0.264 & 0.267 & 0.938 & 0.619 & 0.284 & 0.286 & 0.938 & 0.459 & 0.066 & 0.063 & 0.953 \\
marginal-o0c1 & 1   & 0.059 & 0.136 & 0.137 & 0.930 & 0.200 & 0.144 & 0.146 & 0.936 & 2.663 & 0.044 & 0.043 & 0.952 \\
              & 1.5 & 0.099 & 0.203 & 0.205 & 0.934 & 0.247 & 0.217 & 0.219 & 0.937 & 2.124 & 0.068 & 0.066 & 0.954 \\
              & 2   & 0.148 & 0.268 & 0.270 & 0.932 & 0.285 & 0.287 & 0.290 & 0.940 & 1.575 & 0.092 & 0.089 & 0.952 \\
marginal-o0c0 & 1   & 0.091 & 0.136 & 0.137 & 0.931 & 0.223 & 0.144 & 0.145 & 0.936 & 2.439 & 0.044 & 0.043 & 0.951 \\
              & 1.5 & 0.150 & 0.203 & 0.205 & 0.935 & 0.288 & 0.217 & 0.219 & 0.937 & 1.928 & 0.068 & 0.066 & 0.952 \\
              & 2   & 0.209 & 0.268 & 0.271 & 0.936 & 0.342 & 0.287 & 0.289 & 0.937 & 1.467 & 0.092 & 0.089 & 0.951 \\
\addlinespace[3pt]
\multicolumn{14}{c}{\textbf{Doubly robust estimator (Frailty Cox)}} \\
\addlinespace[2pt]
frailty-o1c1  & 1   & 0.977 & 0.133 & 0.129 & 0.940 & 1.229 & 0.143 & 0.141 & 0.942 & 3.872 & 0.031 & 0.029 & 0.951 \\
              & 1.5 & 1.033 & 0.199 & 0.193 & 0.939 & 1.267 & 0.215 & 0.212 & 0.945 & 2.485 & 0.048 & 0.046 & 0.955 \\
              & 2   & 1.074 & 0.263 & 0.255 & 0.938 & 1.280 & 0.285 & 0.281 & 0.943 & 1.526 & 0.067 & 0.064 & 0.958 \\
frailty-o1c0  & 1   & 1.117 & 0.133 & 0.129 & 0.940 & 1.334 & 0.143 & 0.141 & 0.939 & 3.064 & 0.031 & 0.029 & 0.952 \\
              & 1.5 & 1.242 & 0.199 & 0.193 & 0.936 & 1.433 & 0.215 & 0.213 & 0.940 & 1.634 & 0.049 & 0.046 & 0.953 \\
              & 2   & 1.350 & 0.263 & 0.255 & 0.937 & 1.509 & 0.284 & 0.282 & 0.939 & 0.654 & 0.067 & 0.064 & 0.955 \\
frailty-o0c1  & 1   & 0.775 & 0.137 & 0.132 & 0.944 & 0.977 & 0.146 & 0.143 & 0.946 & 3.121 & 0.060 & 0.053 & 0.965 \\
              & 1.5 & 0.828 & 0.204 & 0.197 & 0.945 & 1.048 & 0.219 & 0.216 & 0.945 & 2.464 & 0.091 & 0.081 & 0.964 \\
              & 2   & 0.873 & 0.270 & 0.261 & 0.943 & 1.097 & 0.290 & 0.286 & 0.945 & 1.950 & 0.123 & 0.108 & 0.963 \\
frailty-o0c0  & 1   & 1.004 & 0.136 & 0.131 & 0.945 & 1.145 & 0.145 & 0.142 & 0.944 & 1.714 & 0.060 & 0.053 & 0.963 \\
              & 1.5 & 1.185 & 0.204 & 0.196 & 0.943 & 1.324 & 0.218 & 0.214 & 0.941 & 0.903 & 0.093 & 0.081 & 0.963 \\
              & 2   & 1.348 & 0.269 & 0.259 & 0.946 & 1.481 & 0.289 & 0.284 & 0.941 & 0.318 & 0.125 & 0.109 & 0.963 \\
\addlinespace[3pt]
\multicolumn{14}{c}{\textbf{Non-parametric RMT-IF}} \\
\addlinespace[2pt]
RMT-IF        & 1   & 2.043 & 0.139 & 0.136 & 0.934 & 2.091 & 0.159 & 0.154 & 0.944 & 1.135 & 0.109 & 0.106 & 0.953 \\
              & 1.5 & 2.900 & 0.206 & 0.202 & 0.931 & 2.901 & 0.239 & 0.230 & 0.937 & 2.898 & 0.165 & 0.161 & 0.948 \\
              & 2   & 3.735 & 0.271 & 0.265 & 0.920 & 3.692 & 0.314 & 0.303 & 0.935 & 4.270 & 0.221 & 0.215 & 0.950 \\
\bottomrule
\end{tabular}%
}
\end{table}

\section{Cluster-level RMT-IF estimates in STRIDE trials}

\begin{figure}
    \centering
    \includegraphics[width=1\linewidth]{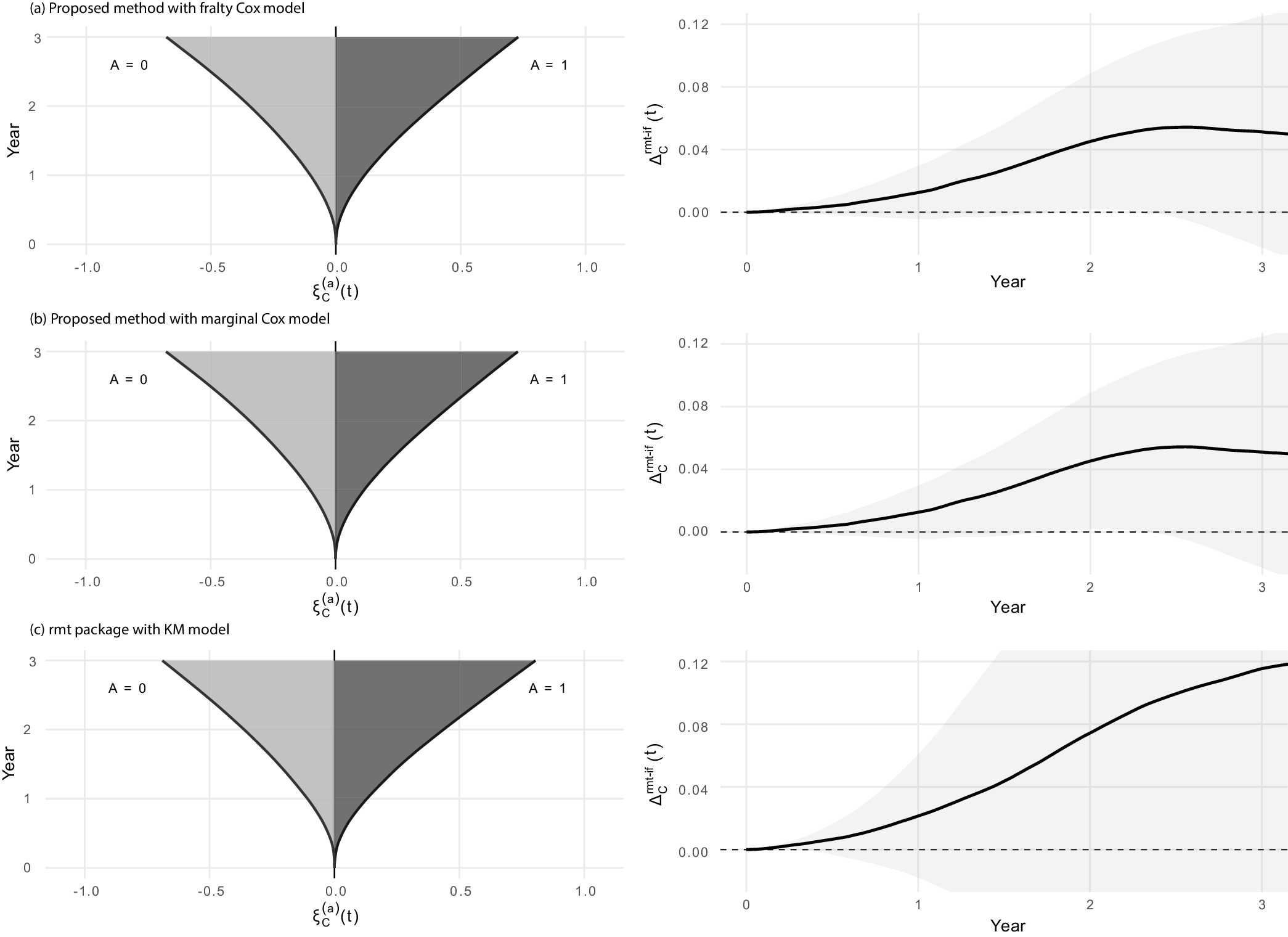}
    \caption{Estimated cluster-level RMT-IF, $\xi_C^{(a)}(t)$ (left panels), and the causal effect, $\Delta_C^{\text{rmt-if}}(t)$ (right panels), from the SPRIDE data. For each method, the left panel displays the bouquet plot of the arm-specific cluster-level RMT-IF, and the right panel shows the corresponding estimated causal effect. Panel (a) reports results from the proposed doubly robust method using stage-specific frailty Cox models for $P\{T_i^{q,(a)} > t \mid \bm V_i\}$ and for the censoring distribution $K_c^{(a)}(t \mid \bm V_i)$. Panel (b) reports results from the proposed doubly robust method using stage-specific marginal Cox models for $P\{T_i^{q,(a)} > t \mid \bm V_i\}$ and for the censoring distribution. Panel (c) reports the nonparametric estimator of Mao \cite{mao2023restricted} using Kaplan--Meier estimators for $P\{T_i^{q,(a)} > t \mid \bm V_i\}$ and $K_c^{(a)}(t \mid \bm V_i)$ with weights $1/N_i$ applied at the cluster level.}
\label{fig:stride_c}
\end{figure}

Figure \ref{fig:stride_c} presents the cluster-level arm-specific RMT-IF estimates and their causal effect from the STRIDE trial, comparing the proposed doubly robust estimators with the nonparametric RMT-IF estimator of Mao \cite{mao2023restricted}. The left panel shows the treatment-specific trajectories of $\xi_C^{(a)}(t)$, and the right panel displays the corresponding contrast $\Delta_C^{\text{rmt-if}}(t)$, each with pointwise 95\% confidence bands. At the cluster level, the two proposed methods, based on different working models, produce very similar trajectories for both $\xi_C^{(a)}(t)$ and $\Delta_C^{\text{rmt-if}}(t)$, and the associated confidence bands include zero over the follow-up period, indicating that the estimated treatment effect is not statistically significant at cluster level. In contrast, the nonparametric estimator exhibits a noticeably different trajectory for the treatment effect.

\section{Distribution of non-fatal events in SRPINT and STRIDE trial} \label{supp:trial_dist}

Tables~\ref{tab:recurrent_nonfatal} and~\ref{tab:recurrent_fatal} summarize the distribution of recurrent event counts by randomized treatment arm for two trial datasets. In the SPRINT trial (Table~\ref{tab:recurrent_nonfatal}), the majority of participants experienced no non-fatal events during follow-up, and only a small fraction of participants experienced multiple non-fatal events, with fewer than 2\% in either arm having two or more events and fewer than 0.1\% having five or more events.  The STRIDE trial self-reported fall injuries (Table~\ref{tab:recurrent_fatal}) have more recurrence. Approximately half of participants reported no fall injuries , while the remainder experienced one or more injuries, with fewer than 1\% of participants in either arm reporting nine or more injuries.

\begin{table}[!htbp]
\centering
\caption{Distribution of non-fatal events by treatment arm in SPRINT Trial.}
\label{tab:recurrent_nonfatal}
\begin{tabular}{lcc}
\toprule
\textbf{Number of recurrent non-fatal events} & \textbf{Control (N = 4616)} (\%) & \textbf{Intervention (N = 4628)} (\%) \\
\hline
0   & 4270 (92.50) & 4368 (94.38) \\
1   & 272 (5.89)   & 202 (4.36)   \\
2   & 55 (1.19)    & 42 (0.91)    \\
3   & 13 (0.28)    & 10 (0.22)    \\
4   & 2 (0.04)     & 4 (0.09)     \\
5+  & 4 (0.09)     & 2 (0.04)     \\
\bottomrule
\end{tabular}
\end{table}

\begin{table}[!htbp]
\centering
\caption{Distribution of recurrent self-reported fall injuries by treatment arm in STRIDE trial}
\label{tab:recurrent_fatal}
\begin{tabular}{lcc}
\toprule
\textbf{Number of self-reported fall injuries} & \textbf{Control (N = 2649)} (\%) & \textbf{Intervention (N = 2802)} \(\%\) \\
\hline
0  & 1412 (53.30) & 1593 (56.85) \\
1  & 695 (26.24)  & 684 (24.41)  \\
2  & 311 (11.74)  & 289 (10.31)  \\
3  & 127 (4.79)   & 129 (4.60)   \\
4  & 51 (1.93)    & 55 (1.96)    \\
5  & 26 (0.98)    & 23 (0.82)    \\
6  & 7 (0.26)     & 11 (0.39)    \\
7  & 9 (0.34)     & 4 (0.14)     \\
8  & 3 (0.11)     & 4 (0.14)     \\
9+ & 8 (0.30)     & 10 (0.36)    \\
\bottomrule
\end{tabular}
\end{table}

\clearpage
\section{Example code for the \texttt{DRsurv} package}
We illustrate \texttt{DRsurvfit()} in the multi-state setting using the example dataset \texttt{datm}. The function has the following declaration:
\begin{verbatim}
DRsurvfit <- function(data,
                      formula,
                      cens_formula = NULL,
                      intv,
                      method   = c("marginal", "frailty"),
                      estimand = c("SPCE", "RMTIF"),
                      trt_prob = NULL,
                      variance = c("none","jackknife"),
                      fit_controls = NULL)
\end{verbatim}
with arguments:
\begin{itemize}
  \item \texttt{data}: A \texttt{data.frame} of individual-level observations.
  \item \texttt{formula}: Outcome model of the form \texttt{Surv(time, event) \textasciitilde\ covariates + cluster(clusterID)}.
  \item \texttt{cens\_formula}: Optional censoring model. If \texttt{NULL}, the RHS of \texttt{formula} is reused.
        \texttt{Surv(time, event == 0)}.
  \item \texttt{intv}: Name of the cluster-level treatment column (0/1), constant within cluster.
  \item  \texttt{id\_var}: name of the individual identifier.
  \item \texttt{method}: \texttt{``marginal''} (Cox PH via \texttt{survival::coxph}) or \texttt{``frailty''} (gamma frailty via \texttt{frailtyEM::emfrail}).
  \item \texttt{estimand}: \texttt{``SPCE''} (survival probability causal effects) or \texttt{``RMTIF''} for RMT-IF estimand.
  \item \texttt{variance}: \texttt{``none''} or \texttt{``jackknife''} for leave-one-cluster-out (LOCO) standard errors.
  \item \texttt{fit\_controls}: Optional \texttt{frailtyEM::emfrail\_control()} list (used only for \texttt{method = ``frailty''}).
\end{itemize}
Here the event indicator takes values in \(\{0,1,2,3\}\), where 0 denotes censoring or remaining at risk, and 1--3 correspond to ordered event types in a prioritized composite. We fit marginal Cox models for both outcome and censoring, compute leave-one-cluster-out (LOCO) jackknife variances, and summarize
\begin{enumerate}
  \item state-specific SPCE at \(t \in \{1,2,3\}\), and
  \item the RMT-IF estimators up to \(\tau \in \{1,2,3\}\),
\end{enumerate}
using two-sided $t$-intervals (\(\alpha = 0.05\), \(\mathrm{df} = M - 1\)).

\begin{verbatim}
library(DRsurvCRT)
data(datm)

fit_rmtif <- DRsurvfit(
  data     = datm,
  formula  = Surv(time, event) ~ W1 + W2 + Z1 + Z2 + cluster(cluster),
  intv     = "trt",
  method   = "marginal",
  estimand = "RMTIF",
  variance = "jackknife"
)

summary(
  fit_rmtif,
  tau    = c(1, 2, 3),
  level  = "cluster",   # cluster-level estimands
  type   = "RMTIF",      # SPCE and RMT-IF
  digits = 3,
  alpha  = 0.05
)
\end{verbatim}

Example output:
\begin{verbatim}
DRsurvfit: method = marginal, estimand = RMTIF
Treatment probs (p0, p1): 0.5, 0.5
Outcome model:   Surv(time, event) ~ W1 + W2 + Z1 + Z2 + cluster(cluster)
Censoring model: Surv(time, event == 0) ~ W1 + W2 + Z1 + Z2 + cluster(cluster)
Cluster id col:  cluster
Clusters (M):    26
Obs (N):         886
Events (status != 0): 448

RMT-IF summary (cluster-level)
        R1 (LCL, UCL)         R0 (LCL, UCL)           R1-R0 (LCL, UCL)       
t=0.996 0.119 (0.0501, 0.188) 0.0772 (0.00214, 0.152) 0.0419 (-0.0181, 0.102)
t=1.99  0.303 (0.151, 0.456)  0.221 (0.0802, 0.362)   0.0824 (-0.0414, 0.206)
t=2.99  0.517 (0.276, 0.759)  0.387 (0.188, 0.587)    0.13 (-0.0628, 0.323)  
  t-intervals with df = 25, alpha = 0.050
\end{verbatim}

\end{document}